\documentclass[12pt]{article}
\usepackage{graphicx,amsmath,amssymb,cite}
\usepackage{epsfig,epsf}
\usepackage{graphicx}
\usepackage{epstopdf}
\newcommand{\beq}{\begin{equation}}
\newcommand{\eeq}{\end{equation}}
\newcommand{\alphabar}{\bar{\alpha_s}}
\newcommand{\betabar}{\bar{\beta_0}}

\numberwithin{equation}{section}

\begin{document}

\begin{flushright}
DESY 15-152\\
\end{flushright}
\vspace*{0.5 cm}

\begin{center}

{\Large{\bf The Behaviour  of the Green Function for the BFKL Pomeron
 with Running Coupling. }}

\vspace*{1 cm}

{\large H. Kowalski~$^1$, L.N. Lipatov~$^{2}$, and D.A. Ross~$^3$} \\ [0.5cm]
{\it $^1$ Deutsches Elektronen-Synchrotron DESY, D-22607 Hamburg, Germany}\\[0.1cm]
{\it $^2$  St.Petersburg State University and  Petersburg Nuclear Physics Institute, Gatchina 188300, St. Petersburg, Russia
}\\[0.1cm]
{\it $^3$ School of Physics and Astronomy, University of Southampton,\\Highfield, Southampton SO17 1BJ, UK}\\[0.1cm]
 \end{center}

\vspace*{3 cm}

\begin{center}
{\bf Abstract}  \end{center}
We analyse here in LO the  physical properties of the Green function solution for the BFKL equation. We show that the solution obeys the orthonormality conditions in the physical region and fulfills the  completeness requirements. The unintegrated gluon density is shown to consists of a set of few poles with parameters  which could be determined by comparison with the DIS data of high precision.


\vspace*{3 cm}


\newpage

\section{Introduction}
The BFKL equation is particularly well suited for description of the behaviour of gluon density in the low-$x$ region. Its application in this region is of major importance for the LHC and cosmic ray physics. In recent years we have therefore investigated~\cite{KLRW,KLR}  the solution of this equation using the discrete eigenfunction method, first proposed in~\cite{lipatov86}. The method is closely connected to the Green function approach which is, in our view, the most suitable since it does not require any cutoff on the BFKL dynamics.    

The results reported in these papers where as interesting as they were puzzling. The proper description of data was only achieved by involving a large number of eigenfunction,  O(100),  contributing in a slowly convergent way.  On the other hand,  the third and higher eigenfunctions were already  sensitive to physics at very large scales, much beyond the $Q^2$ region of data. If true, this property would offer a  framework for investigations of Beyond the Standard Model (BSM)
physics at large energy scales using relatively moderate beam energies combined with precision measurements. 

The slow convergence of the procedure given in~\cite{KLRW,KLR}  suggests,
in particular,  that  the discrete eigenfunctions might not by themselves form a complete set.
Therefore, in a recent paper~\cite{LKR1}  we rederived the  BFKL Green function using the complete BFKL spectrum and showed 
 how the imposition of both UV and IR boundary conditions leads to a discrete set of poles. 
The purpose of this paper is to investigate the properties of the full Green function solution and to relate it to our previous work.

 The paper is organized as follows: In section \ref{sec1} we recapitulate  the results of~\cite{LKR1} which were obtained in the semi-classical approximation. To illustrate the role of approximations used in our method  we discuss
  in section \ref{sec2}  
the case in which the coupling runs without thresholds. In this case it 
is possible to obtain the Green function in  analytic form, which reduces
the solution to an expression in terms of Airy functions in the diffusion approximation
 in which the characteristic function, $\chi(\nu)$,
 is simplified to  a quadratic function of oscillation frequency, $\nu$. In  section \ref{sec3}
 we discuss a restriction that is imposed on  possible paths for the
integration over the Mellin transform variable, $\omega$, arising from
 thresholds in the running of the coupling, and discuss the agreement of 
 numerical results using two substantially different $\omega$-paths.
 In section \ref{sec4} we discuss the orthonormality and completeness
 of the BFKL eigenfunctions  obtained within the approximation
that we are using and report that in order to obtain the required
 completeness relation it is necessary to include the continuum of states 
 for which $\omega$ is negative.
In section \ref{sec5} we discuss the behaviour of the residues of the poles 
as the gluon transverse momentum increases and show that the leading pole
 is attenuated so that the 
sub-leading poles acquire an ever-increasing significance as the transverse momentum increases.
 In section \ref{sec6} we present the results for the unintegrated gluon
density in a model in which a very simple ansatz is used for the
 proton impact factor and for the phase of the oscillations
 of the wavefunctions in the infrared
regime.
 In section \ref{sec7} we discuss
 the convergence of the sum over poles and show that, in contrast to the results of ref.~\cite{KLRW,KLR},  
in the formalism that we are using here this convergence is quite rapid.
 Section \ref{sec8} discusses the role of the continuum
 for negative $\omega$ on the calculation of the unintegrated gluon density and section 7.3 shows the relation to the DGLAP calculation.
 In section 7.4 we discuss the behaviour of the gluon density as a function of $x$ at fixed low virtualities, $k_T^2$,  and show that qualitatively it has properties similar to data and display clear contributions from  subleading poles.  
In section \ref{sec9} we show how the presence of new physics albeit at very
 large energies affects the running of the coupling such that both the
 positions and the residues of the poles are altered and this in turn
 gives rise to a measurable change in the unintegrated gluon density.
 Section \ref{sec10} contains our conclusions.


\section{Green Function of the BFKL equation} 
\label{sec1}
We showed~\cite{LKR1} that the 
 (Mellin transform of the) Green function for the
BFKL equation with running coupling can be solved in the semi-classical
 approximation in terms of Airy functions, $Ai(z)$ and $Bi(z)$, where $z$
 is a function of the transverse momentum of the QCD pomeron, $k_T$, and the Mellin
 transform variable, $\omega$. 

More precisely the equation for the Green function  is
\beq \omega {\cal G}_\omega(t,t^\prime)
 -  \int dt^{\prime\prime} \sqrt{\alphabar(t)}
 {\cal K}(t,t^{\prime\prime}) \sqrt{\alphabar(t^{\prime\prime})}
 {\cal G}_\omega(t^{\prime\prime},t^\prime) = \delta(t-t^\prime), 
 \label{gf1} \eeq
where $t=\ln(k_T^2/\Lambda^2)$.
We note that we have introduced the running of the coupling in such a
way that the hermiticity of the kernel is preserved so that its eigenfunctions
 form a complete  orthonormal set.
 Eq.(\ref{gf1})  has a solution,
in the semi-classical approximation, of the form 
\beq {\cal G}_\omega(t,t^\prime) \ = \ \pi 
  N_\omega(t)N_\omega(t^\prime)
 \left[ Ai(z(t)) Bi(z(t^\prime) \theta(t-t^\prime) + t \leftrightarrow t^\prime
 \right]   \label{sol1} .\eeq 
The argument $z(t)$ of the Airy functions is given by
 \beq \left(-z(t) \right)^{3/2} \ \equiv \  \frac{3}{2} \int_t^{t_c} dx \nu_\omega(x),  \label{arg} \eeq
where  
 \beq \chi\left(\nu_\omega(t)\right) \ =  \ \frac{\omega}{\alphabar(t)} 
 \label{freq} \eeq 
and $\chi(\nu)$ are the eigenvalues of the kernel ${\cal K}$.
The parameter $t_c$ is the ($\omega$-dependent) 
value of $t$ for which $\nu_\omega(t_c)=0$.
 $N_\omega(t)$ 
is a normalization factor given by
 \beq N_\omega(t) \ = \
 \frac{|z(t)|^{1/4}}{\sqrt{\frac{1}{2} \alphabar(t) \chi^\prime\left(\nu_\omega(t) \right)}}
\label{norm1} \eeq

 This Green function is analytic in the entire
 $\omega$-plane with the exception of an essential singularity at $\omega=0$.

The expression on the RHS of  eq.(\ref{sol1}) has the desired ultraviolet
 behaviour, namely it is exponentially attenuated as $t \to \infty$,
 but the infrared behaviour (where $t$ is  small) is {\it not}
 properly  determined.
To obtain the most general solution to (\ref{gf1}) this Green function should be generalized by adding to $Bi$ the solution of the homogeneous BFKL equation,
 i.e. the transformation
 \beq Bi(z(t)) \ \to \overline{Bi}(z(t)) \ \equiv \ Bi(z(t))
 + \cot\left(\phi(\omega)\right)Ai(z(t)). 
 \label{transform}, \eeq
and eq.(\ref{sol1}) becomes
\beq {\cal G}_\omega(t,t^\prime) \ = \ \pi 
  N_\omega(t)N_\omega(t^\prime)
 \left[ Ai(z(t)) \overline{Bi}(z(t^\prime) \theta(t-t^\prime) + t \leftrightarrow t^\prime
 \right]   \label{sol1a} .\eeq 

 The transformation
 (\ref{transform}) plays the dual role of providing a set of poles
 at the values of $\omega$ for which the function $\phi(\omega)=n\pi$
 and fixing the phase of the oscillatory behaviour of the Green function
 for very small $t$ (or $t^\prime$), thereby providing the connection
 between the determination of the infrared behaviour of the Green function
 and the position of the poles, originally suggested in \cite{lipatov86}.

It was pointed out in \cite{LKR1} that upon inverting the Mellin transform
 of the amplitude, by integrating along a suitable path  in the $\omega$-plane, 
the saddle-point approximation, used to match the BFKL solution  with the result of a DGLAP
 analysis \cite{ccs,tw,abf}, could be valid provided the saddle-point was to the right of
 all the poles in the $\omega$-plane. On the other hand, if this is not the case
then the selected contour for the integral over $\omega$ must surround
 the poles to the right of the saddle-point and will provide significant
 supplementary contributions to the unintegrated gluon density which are not
 matched to the DGLAP result.

In this paper we report on a numerical analysis of the above-mentioned
 semi-classical  solution and discuss the behaviour of the eigenfunctions
 of the kernel. We confine ourselves to the leading
 order BFKL kernel. The effects of the large components of the NLO
 BFKL kernel will be discussed in a subsequent paper.

\section{Explicit Solution in the Absence of Thresholds}
\label{sec2}
We begin by considering the simplified case in which the running
 coupling is given by
\beq \alphabar(t) \ = \ \frac{1}{\betabar{t}} \eeq
and $\betabar$ is a constant.

Consider the eigenfunctions, $f_\omega(t)$ of the kernel with running
coupling
\beq \int dt^\prime \sqrt{\alphabar(t)}
 {\cal K}(t,t^\prime)\sqrt{\alphabar(t^\prime)}
 f(\omega,t^\prime) \ = \ \omega f_\omega(t), \label{ft} \eeq

 The eigenfunctions may be written in integral form as
 \beq f_\omega(t) \ = \ \sqrt{\frac{t}{2\pi\omega}} \int_{\cal C} 
d\nu g_\omega(\nu) e^{i\nu t}  \label{gen1} \eeq
with
\beq \ln \left(g_\omega(\nu)\right)  \ = \ \frac{-i}{\betabar\omega}
\int_0^\nu \chi(\nu^\prime) d\nu^\prime .  \eeq
Provided $g_\omega(\nu)$ is square-integrable,
the contour ${\cal C}$ 
 may be taken to be along the real axis
in the $\nu$-plane  and we  have
 \beq f_\omega(t) \ = \ \sqrt{\frac{t}{2\pi\omega}} \int_{-\infty}^{\infty} 
d\nu g_\omega(\nu) e^{i\nu t}  . \eeq


These eigenfunctions obey the orthonormality relation
\beq  \int dt f_\omega(t) f^*_{\omega^\prime}(t) 
\ = \ 2\pi \delta(\omega-\omega^\prime) \eeq 
and completeness relation
\beq \int d\omega f_\omega(t) f^*_\omega(t^\prime) \ = \ 2 \pi \delta(t-t^\prime)
\eeq

For the leading order BFKL equation with running coupling, 
$g_\omega$ is given by
\beq g_\omega(\nu) \ = 
 g^*_{\omega}(-\nu) \ = \ \left( g^*_\omega(\nu) \right)^{-1}  \ = \  
\ e^{2i \gamma_E \nu/(\betabar \omega)}
 \left[\frac{\Gamma\left(\frac{1}{2}+i\nu\right)}{\Gamma\left(\frac{1}{2}-i\nu\right)}
   \right]^{1/(\betabar \omega)}. \eeq

The integral over $\nu$   in eq.(\ref{gen1})
 can be approximated
by a Gaussian integral around the saddle-point, $\nu_\omega(t)$, given by
 \beq \chi\left(\nu_\omega(t)\right) \ = \ \betabar \omega t  \eeq
For sufficiently small $t$  the eigenfunctions have an oscillatory behaviour
\beq f_\omega(t) \propto 
\sqrt{\frac{t}{\chi^\prime\left(\nu_\omega(t)\right)}}
 \cos\left(\frac{\pi}{4}+ t \nu_\omega(t) - \frac{1}{\betabar \omega}
 \int_0^{\nu_\omega(t)} d\nu^\prime \chi(\nu^\prime)\right) \eeq

In terms of the  function $g_\omega(\nu)$, the Green function
may be written (see \cite{Lipatov14})  as
\beq {\cal G}_\omega(t,t^\prime) \ = \ -\frac{\sqrt{tt^\prime}}{4\pi\omega}
 \int_{-\infty}^\infty d\nu \int_{-\infty}^\infty d\nu^\prime e^{i t \nu}
e^{i t^\prime \nu^\prime} \left[i\epsilon(\nu+\nu^\prime)+ c(\omega)
 \right]  g_\omega(\nu)g_\omega(\nu^\prime) \label{gftsol1}, \eeq
where $c(\omega)$ is an arbitrary function of $\omega$ and this second term
 reflects the fact that one can add to a Green function any solution 
 to the homogeneous part of the equation for the Green function.
If we write
 \beq c(\omega) \ \equiv \cot\left(\phi(\omega)\right), \eeq
then in the semi-classical limit, we can interpret $\phi(\omega)$ as being
 the  phase of the oscillations of the BFKL eigenfunctions at some
 small value of $t$.

That (\ref{gftsol1})  is indeed the solution
to the equation for the Green function  can be seen by applying 
 the (hermitian) operator
$$ {\cal O}(t,t^\prime) \ \equiv \ 
\omega \delta(t-t^\prime)-\frac{1}{\sqrt{\betabar t}}
 {\cal K}(t,t^\prime) \frac{1}{\sqrt{\betabar t^\prime}} $$
and using
\beq \int dt^\prime {\cal K}(t,t^\prime) g_\omega(\nu) e^{i\nu t^\prime}
 \ = \ - i \betabar \omega \frac{d g_\omega(\nu)}{d\nu} e^{i\nu t} \eeq
to get
\begin{eqnarray} \int d\tau  {\cal O}(t,\tau)  {\cal G}_\omega(\tau,t^\prime)
  & = & - \frac{\sqrt{tt^\prime}}{4\pi}
 \int_{-\infty}^\infty d\nu \int_{-\infty}^\infty d\nu^\prime \left\{
 \left(  g_\omega(\nu)  + \frac{i}{t} \frac{d g_\omega(\nu)}{d\nu}  \right)
 \left[i\epsilon(\nu+\nu^\prime)+\cot\left(\phi(\omega)\right)
 \right]  \right.  \nonumber \\ & & \left. \times 
 g_\omega(\nu^\prime)   e^{i t \nu}
e^{i t^\prime \nu^\prime} \right\} \end{eqnarray}

Integrating over $\nu$ by parts 
and using $g_\omega(-\nu)=(g_\omega(\nu))^{-1}$
yields
 \beq \int d\tau  {\cal O}(t,\tau)  {\cal G}_\omega(\tau,t^\prime)
  \ = \  \frac{\sqrt{tt^\prime}}{4\pi}
\int_{-\infty}^\infty d\nu \int_{-\infty}^\infty d\nu^\prime
 2\delta(\nu+\nu^\prime)  e^{i t \nu}
e^{i t^\prime \nu^\prime} g_\omega(\nu) g_\omega(\nu^\prime)  \ = \ \delta(t-t^\prime).  \eeq

Thus we see that it is the  factor $\epsilon(\nu+\nu^\prime)$
 inside the integration over $\nu$ and $\nu^\prime$ that generates
the required inhomogeneous term in the Green-function equation.

For small $t$ ($t < t^\prime$) this Green function has an oscillatory $t$ behaviour
\beq {\cal G}_\omega(t,t^\prime) \sim \sin\left( \phi(\omega)+ 
\frac{\pi}{4}+ t \nu_\omega(t) - \frac{1}{\betabar \omega}
 \int_0^{\nu_\omega(t)} d\nu^\prime \chi(\nu^\prime)\right) . \label{greenosc} \eeq
The phase of these oscillations fixed by the boundary conditions
 at some small value, $t_0$, of $t$ determines
 the function $\phi(\omega)$ and hence the positions of the poles at $\omega=\omega_n$
 where $\phi(\omega) \, = \,n\pi$.

The Green function has a spectral representation in terms of these poles, namely
\beq {\cal G}_\omega(t,t^\prime) \ = \ \sum_n
 \frac{f_{\omega_n}(t)f_{\omega_n}(t^\prime)}{(\omega-\omega_n)} 
 \label{spectral} \eeq
and we see from the completeness relation that
 \beq \lim_{\omega \to \infty} {\cal G}_\omega(t,t^\prime) \ = \ \frac{\delta(t-t^\prime)}{\omega}
 \label{asymp} . \eeq

In the limit $t^\prime \, \to \, -\infty$ (keeping $t$ fixed), 
the integral over $\nu^\prime$ may be approximated by its contributions in the
regions of the two saddle points at
 $$ \nu^\prime \ =  \ \pm\nu_\omega(t^\prime) $$
where
 \beq
 \chi(\nu_\omega(t)) \ = \ \betabar \omega t. 
 \label{nuom}\eeq  
In this limit, $|\nu_\omega(t^\prime)| > |\nu|$ and so the function
 $\epsilon(\nu+\nu^\prime)$ is replaced by $\pm 1$ at the saddle points
 $\pm\nu_\omega(t^\prime) $ respectively.

 We define
\beq s_\omega(t) \ \equiv \ t \nu_\omega(t)-\frac{1}{\betabar \omega}
 \int_0^{\nu_\omega(t)} \chi(x) dx .\eeq

Performing the gaussian integrals over $\nu$ and $\nu^\prime$ 
around the two saddle-points we obtain
\begin{eqnarray}
 \lim_{t^\prime\to -\infty} {\cal G}_\omega(t,t^\prime) 
 &=& -\frac{1}{2} \sqrt{\frac{\pi t^\prime}{\chi^\prime(\nu_\omega(t^\prime)}}
\left\{ i 
\left( \frac{e^{is_\omega(t^\prime)}}{\sqrt{i}} -\frac{e^{-is_\omega(t^\prime)}}{\sqrt{-i}} 
 \right) \right. \\ & & \hspace*{1cm} \left.
 +\cot\left(\phi(\omega)\right)
\left( \frac{e^{is_\omega(t^\prime)}}{\sqrt{i}}+\frac{e^{-is_\omega(t^\prime)}}{\sqrt{-i}}  \right)
\right\} f_\omega(t) \nonumber  \\ & \hspace*{-2.1cm} =& \hspace*{-1.3cm}
- \sqrt{\frac{\pi t^\prime}{\chi^\prime(\nu_\omega(t^\prime)}}
 \left\{ \cos\left(s_\omega(t^\prime)+\frac{\pi}{4}\right)
+\cot\left(\phi(\omega)\right) \sin\left(s_\omega(t^\prime)+\frac{\pi}{4}
 \right)
\right\} f_\omega(t) 
\end{eqnarray}

Exploiting the symmetry under $t \, \leftrightarrow \, t^\prime$, we arrive at the semi-classical
 approximation for the Green funcion in the region
 $$t, t^\prime \, \ll \, t_c \ \equiv \frac{4\ln 2}{\betabar \omega} $$

\begin{eqnarray}
 {\cal G}_\omega(t,t^\prime) & = &
\frac{\pi \sqrt{tt^\prime}}{ \sqrt{ |\chi^\prime(\nu_\omega(t)||\chi^\prime(\nu_\omega(t^\prime)|}}
\left\{ 
\theta(t-t^\prime) 
\sin\left(s_\omega(t)+\frac{\pi}{4}\right) \right. \nonumber \\ & &  \hspace*{1cm} \times \left.
  \left[ \cos\left(s_\omega(t)+\frac{\pi}{4}\right)+ \cot(\phi(\omega)) \sin\left(s_\omega(t)+\frac{\pi}{4}\right)
\right]
+ t \leftrightarrow t^\prime \right\}  \label{fme}, \end{eqnarray}

This is the approximation to the Green function given in eq.(\ref{sol1a}) when $t,t^\prime \, \ll \, t_c$. 
A similar argument, using the semi-classical approximation, can be used to
 show that the expression (\ref{sol1a}) matches the semi-classical approximation
 when $t$ or $t^\prime \, \gg \, t_c$
However, the semi-classical
 approximation breaks down if $t \ \mathrm{or} \ t^\prime \, \approx t_c$, since $\nu_\omega(t)$ becomes very small and the
 curvature at the saddle-points becomes small. 
Nevertheless, as we have shown in \cite{LKR1}, in this limit the characteristic 
function may be approximated by a function which is quadratic in $\nu$
such that the homogeneous part of the equation for the Green function
 reduces to Airy's equation and the particular combination of Airy functions
 given in eq.(\ref{sol1a}) generates the appropriate inhomogeneous
 term, so that the expression (\ref{sol1a}) is a good approximation
 to  the Green function in all regions of $t$ and $t^\prime$.

Henceforth we take the running coupling to be given by
\beq \frac{1}{\alphabar(t)}=\frac{1}{\alphabar(t_0)}+\int_{t_0}^t \betabar(t^\prime) dt^\prime
 \label{running} \eeq
where $\betabar(t)$ has steps at the heavy particle thresholds.
With this more realistic function for the running coupling, it is no longer possible
to solve the Green function analytically, even in integral form, and we confine
ourselves to a numerical analysis within the semi-classical approximation for which 
the integral over the variable $\nu$ has a saddle-point at $\nu_\omega(t)$.

\section{Numerical Solution Using Different Paths on the $\omega$-Plane}
\label{sec3}
The unintegrated gluon density $\dot{g}(x,t)$ is given by the inverse Mellin transform
 of the Green function by
\beq \dot{g}(x,t) \ = \ \frac{e^{t/2}}{2\pi i} \int_{{\cal C}} d\omega x^{-\omega} \int dt^\prime
  {\cal G}_\omega(t,t^\prime) \Phi_P(t^\prime),  \label{gdot} \eeq
where $\Phi_P(t)$ is the proton impact factor. We take the Green function to be given by
eq.(\ref{sol1a})
 and the running coupling given by eq.(\ref{running}). The function
  $\phi(\omega)$ is given by
\beq \phi(\omega) +\frac{\pi}{4} + \int_{t_0}^{t_c} \nu_\omega(t^\prime) dt ^\prime
 = \eta_{NP} \label{phiom} \eeq
with $t_c$ given by the relation
$$ \chi(0) \ = \ \frac{\omega}{\alphabar(t_c)}. $$
This means that at some small value, $t_0$,
 of $t$ the phase of the eigenfunction
with eigenvalue $\omega$  is given by a non-perturbative phase, $\eta_{NP}$.
\footnote{ $\eta_{NP}$ can be a function of $\omega$ but in this 
paper we take it to be constant, although for a realistic fit to data
we would expect it to possess some  $\omega$ dependence.
Note that $\phi(\omega)$ does not depend on the choice of the infrared
 scale, $t_0$, but the infrared phase $\eta_{NP}$ does.}

For a numerical evaluation of the integral over $\omega$, it would be most efficient
 to identify the saddle-point of the integrand and select a path
for $\omega$ which passes through that saddle point and follow the path of steepest descent.
In the case where the saddle point lies to the left of any of the identified poles of
the Green-function the integral must be supplemented by the integral over a contour
surrounding all poles to the right of the saddle-point. \footnote{A full discussion of this is found
in \cite{LKR1}}.

Unfortunately, there are restrictions on the permitted paths in the case where the running
coupling encounters thresholds.
 In order to
 consider complex values of $\omega$, we require complex values of $\alpha_s(t)$ 
and hence complex values of $t$.
In the presence of fermion masses $m_i$, the running coupling to leading order
 is given in terms of its measured value at $t=\overline{t}$ by~\cite{GP}
 \beq \frac{1}{\alphabar(t)}=\frac{1}{\alphabar(\overline{t})}+
 \frac{11}{12} (t-\overline{t})-\frac{1}{9} \sum_i \left[ 
 F\left(\frac{\Lambda^2 e^t}{4m_i^2}\right)
-F\left(\frac{\Lambda^2 e^{\overline{t}}}{4m_i^2}\right)
 \right] \eeq
where the function $F$, given by
$$ F(x)=\sqrt{\frac{(1+x)}{x}} \ln\left(\sqrt{(1+x)}+\sqrt{x}\right), $$
is multi-valued in the complex plane.
The running coupling  is therefore  only uniquely specified in terms of $t$
 corresponding to  $k_T$  covering the complex plane
once so that the  imaginary part of $t$ is restricted to
$$ - \pi \ \leq \ \Im m\{t\} \  < \ \pi, $$ 
which restricts the imaginary part of $\alphabar$ and consequently
the imaginary part of $\omega$. In particular, the calculation of the
argument of the Airy functions requires the identification of 
$\alphabar(t_c)$,
  so that the restriction on the range of the imaginary part
 of $\alphabar$ leads to a corresponding restriction on the imaginary
 part of $\omega$. If the real part of $\omega$ is small then 
the real part of $\alpha(t_c)$  is small, i.e. the real part of $t_c$
 is large. The restriction on the allowed range of the imaginary part
of $t_c$ therefore implies that
the imaginary
part  of $\omega$ must also be small - i.e. we need to select paths which are very close
 to the real axis in this region.
  Furthermore, the restriction on the imaginary part pushes the possible paths towards the essential singularity at very small $\omega$ so that  we could perform the integration only to some small value of  $\omega_{min}$ of around 0.05.

\begin{figure}[htbp]
\centerline{\includegraphics[width=10cm,angle=0]{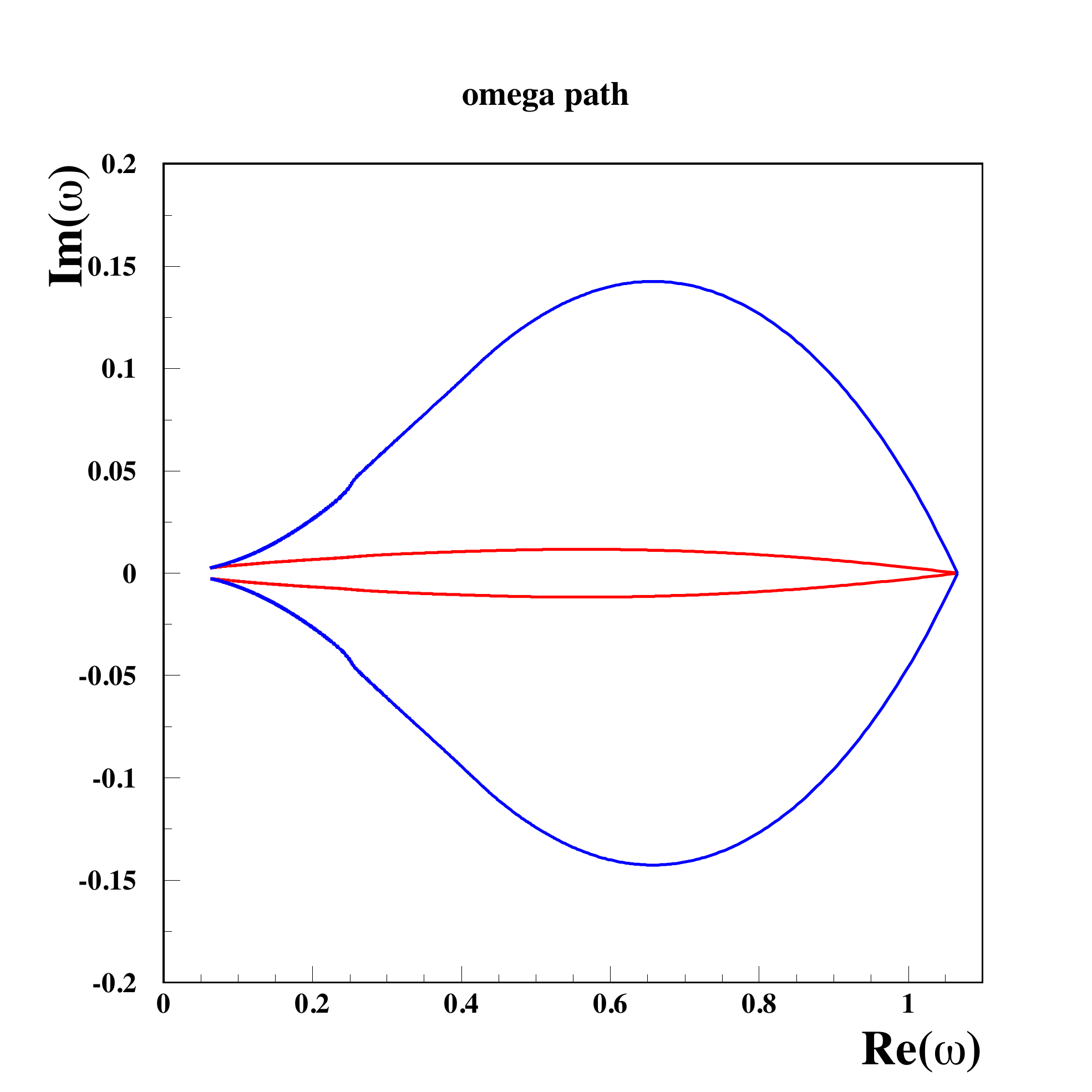}}
\caption{Two different paths in the $\omega$-plane which were used to invert the
 Mellin transform of the Green function.   }
\label{twopaths} \end{figure}

We have selected
 several paths whose imaginary part differ substantially
for large $\omega$. Two of them are shown as an example in Fig.\ref{twopaths}. We have 
performed the integral of eq.(\ref{gdot}) along all of these paths
and find negligible difference over
a large range of $t$ for $t=1$
to $t=17$ (corresponding to  transverse momentum $k_T \approx 2$ TeV).
\footnote{Throughout this paper we have taken
 the QCD scale  $\Lambda$ to be  350 MeV}

In principle, this result should be expected, since there are no singularities 
 of the Green function off the real axis and therefore one can deform the
 contour into any contour that surrounds the real axis and crosses the real axis
 to the right of all poles. However, our contours are not fully closed although the results, at lower $t$ values\footnote{e.g. in the $t$ region corresponding to $k_T < 100$ GeV at $x = 10^{-3} $}, were independent of the  $\omega_{min}$ value. This means that the missing piece of the paths gave a negligible contribution, in this $t$ region. The independence of the results from $\omega_{min}$ was a first sign that the full Green function of  eq.(\ref{gf1}) and eq.(\ref{transform})  behaves differently from the slowly converging sum of eigenfunctions constructed in ref.~\cite{KLRW,KLR}.   In Section \ref{sec7} we will explain in details why this happens.

In addition, the good agreement of the integrals over the different paths  shows the numerical consistency of our Mellin transform integration.

\begin{figure}[h] 
\centerline{\includegraphics[width=9cm]{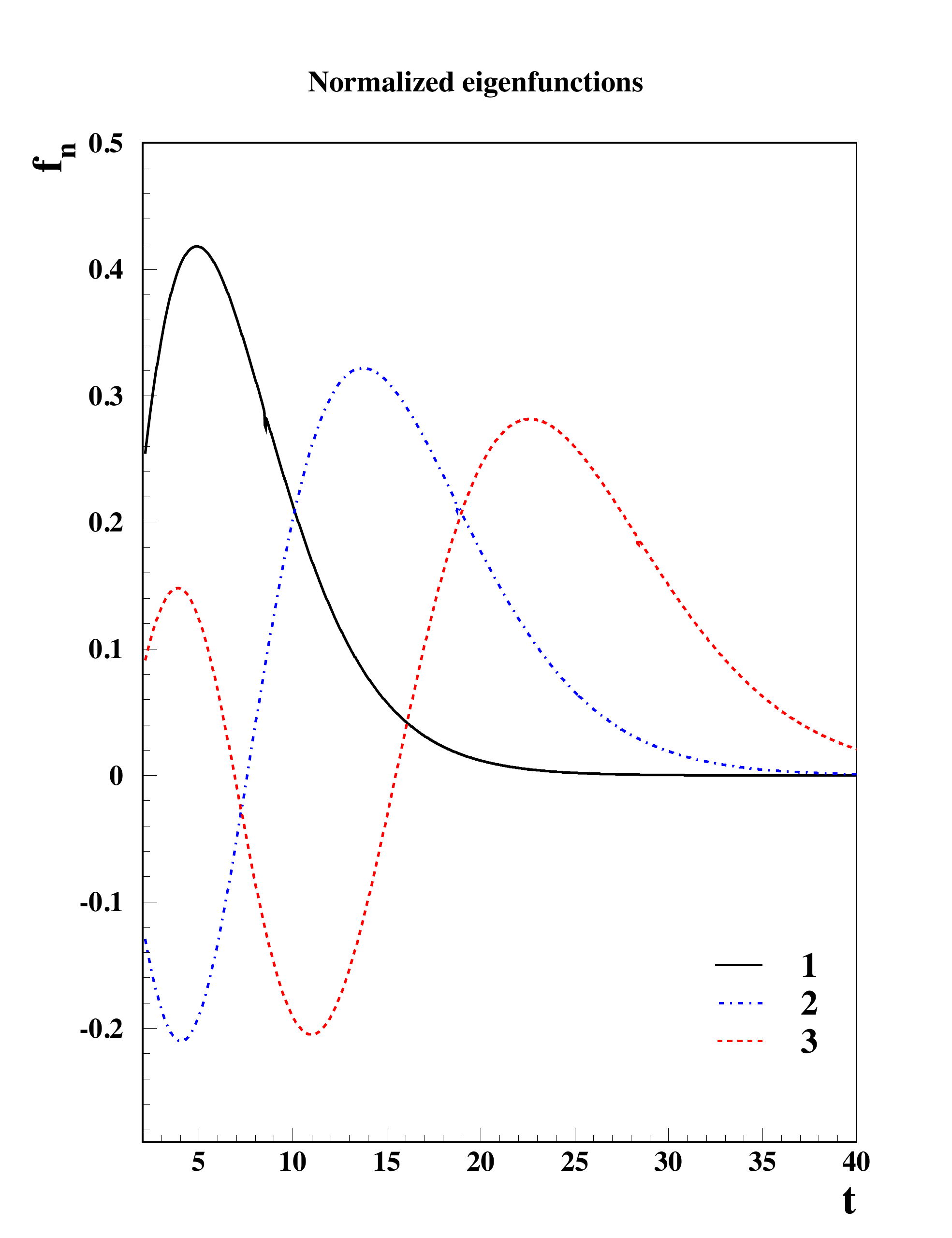} }
\caption{ The first three discrete eigenfunctions, $f_n$  plotted against $t$,
 showing that these eigenfunctions oscillate through $(n-1)$ nodes and are
 then exponentially attenuated.}
\label{fig2} \end{figure}

\section{Properties of the Eigenfunctions}
\label{sec4}

The Green function of the BFKL equation
\beq {\cal G}_\omega(t,t^\prime) \ = \ \pi
  N_\omega(t)N_\omega(t^\prime)
 \left[ Ai(z(t))  \overline{Bi}(z(t^\prime) \theta(t-t^\prime) + t \leftrightarrow t^\prime
 \right]   \label{solfull} .\eeq 
has poles at $\omega=\omega_n$ where $\phi(\omega_n)=n\pi.$
If the infrared phase $\eta_{NP}$ is set to a constant ($\omega$-independent)
value of $-\pi/4$, the first few eigenvalues are shown in Table 1.

\begin{table}[h]
\begin{center}
 \begin{tabular}{||c|c|c|c|c|c|c|c|||} \hline 
  $n$  &$1$& $2$& $3$& $4$& $5$& $6$ & $7$  \\ \hline
  $\omega$ &0.389 & 0.207 & 0.145 & 0.113 & 0.093 & 0.078 & 0.0682 \\
   $t_c$ &8.52 & 18.7 & 29.2 &  38.6 & 48.0 & 57.2 & 66.2 \\
 \hline \end{tabular}
\caption{ The first 7 eigenvalues with the corresponding $t_c$ values for $\eta_{NP}=-\pi/4$}
\end{center}   \end{table}

Except for the first three of these, the eigenvalues are well approximated
 by
 $$ \omega_n \ = \ \frac{1}{1.9(n+\frac{\pi}{4})}, $$ 
consistent with the estimate found in \cite{lipatov86} and \cite{KLRW}.

These correspond to  a discrete set of eigenvalues of the BFKL
 kernel with running coupling whose normalized eigenfunctions are given by
\beq f_{\omega_n}(t) \ = \ \sqrt{\frac{\pi}{\phi^\prime(\omega_n)}} N_{\omega_n}(t)Ai(z(t)).  \label{normeigen} \eeq 
These functions look superficially like the eigenfunctions of ref. \cite{KLRW,KLR}; the Airy functions were defined exactly in the standard way and the
 normalization coefficient was previously 
determined numerically (from the requirement that every eigenfunction should be normalized to unity). Here the normalization coefficient, $N_{\omega_n}$,
is known analytically and is given by eq.(\ref{norm1}); 
the first factor in eq.(\ref{normeigen}) arises from the conversion between the continuum and discrete eigenfunctions. The continuum ones are normalized to   a $\delta$-function in $\omega$ whereas the discrete ones are normalized to a Kronecker  $\delta$-function in the eigenfunction number, $n$, i.e.
\beq \int  f_{\omega_n}(t) f^*_{\omega_{n^\prime}}(t) dt \ = \ \delta_{nn^\prime} \label{orthog} \eeq
This conversion factor can be obtained from the relation
 $$ \int_{\omega_n}^{\omega_n+\Delta \omega_n} \phi^\prime(\omega) d\omega \ = \pi $$
where $\Delta \omega_n$ is the separation of the $n^{th}$ and $(n+1)^{th}$
eigenvalues. For large $n$ the eigenvalues are closely packed and the
separation to the eigenvalues may be approximated by
 $$ \Delta_{\omega_n} \ = \ \frac{\pi}{\phi^\prime(\omega_n)}. $$

The eigenfunctions start by oscillating for small $t$  with $n$
 turning points for $t \, < \, t_c$ and are evanescent for $t \, > \, t_c$. The first 
three such eigenfunctions are plotted in  Fig.\ref{fig2}

\begin{figure}[h]
\centerline{\includegraphics[width=11cm]{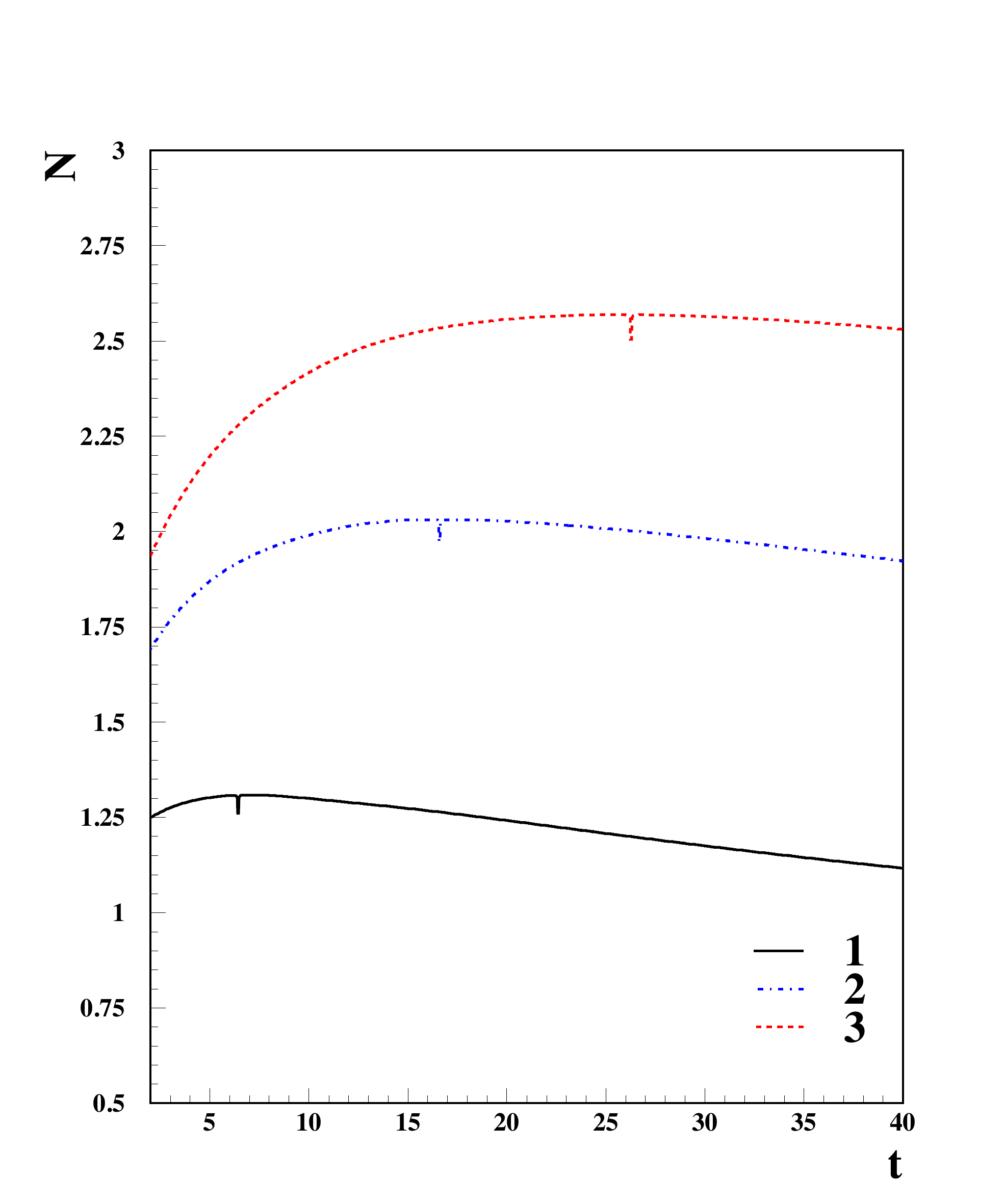} }
\caption{ The normalization factors (\ref{norm1}) for the first three discrete eigenfunctions.}
\label{fig3} \end{figure}

In Table 2 
 we show the numerical results for the  orthonormality relation (\ref{orthog}) evaluated for the first 7 eigenfunctions of  eq.(\ref{normeigen}). 
\begin{table}[h]
\begin{center}
 \begin{tabular}{||c|c|c|c|c|c|c|c|||} \hline 
 {$n\backslash n'$}& $1$& $2$& $3$& $4$ & $5$& $6$ & $7$  \\ \hline
 1 &  1.055   &    -0.0301    &  0.020  &  -0.0158  &  0.0124 &  -0.0107  &   0.0093 \\
 2 &   -0.030 &     1.011      &  -0.008  &  0.007    &  -0.005  &  0.005   & -0.004 \\
 3 &  0.020 &  -0.009  &     1.005  &    -0.005       &  0.004   &  -0.003  &  0.003 \\
 4 &  0.016 &  -0.007  &   -0.005   &    1.003        &    -0.002   &  0.002  &  -0.002 \\
 5 &  0.012 &  -0.005  &    0.004   & -0.002  &    1.001    &    -0.003   &  0.002   \\
 6 &  -0.011 &  0.005  &    -0.003   & 0.002  &    -0.003    &    1.000   &  -0.004   \\
 7 &  0.009 &  -0.004  &     0.003   & -0.002  &    0.002    &    -0.004   &  1.000   \\
 \hline \end{tabular}
\caption{ The orthonormality relation for the first 7 eigenfunctions for $\eta_{NP}=-\pi/4$}
\end{center}   
\label{taborth}
\end{table}

We see that the use of the semi-classical approximation has only
 had a very small effect on the orthonormality relation (\ref{orthog}).
 Furthermore, 
in order to preserve the validity of the perturbative expansion,
 we cannot integrate over all values of $t$, but only 
for $t > t_0$. However, the eigenfunctions are small at small values of $t$.
To see this we note that the additional semi-classical factor, $N_\omega(t)$,  given by eq.(\ref{norm1})
 is constant for $t \approx t_c$, where the Airy functions alone 
is a good approximation to the eigenfunction, but for 
sufficiently small $t$, the variable $z$ and $\chi^\prime\left(\nu_\omega(t)\right)$
 both become very insensitive to $t$ so that  $N_\omega(t)$ then has a $t$ dependence
 $\sim 1/\sqrt{\alphabar(t)}$ - i.e. it decreases in the infrared region as 
$\alphabar(t)$ grows.
The $t$ dependence of these normalization factors for the first three 
eigenfunctions is shown in Fig.\ref{fig3}.

From Table 2 we find that the orthonormality condition is obeyed to a very high accuracy. We consider this as a strong indication that we are using a consistent set of physical  approximations; in particular the semi-classical approximation and the running of $\alpha_s$ are preserving the hermiticity of the Hamiltonian. 



\begin{figure}[htbp]
\centerline{\includegraphics[width=11cm]{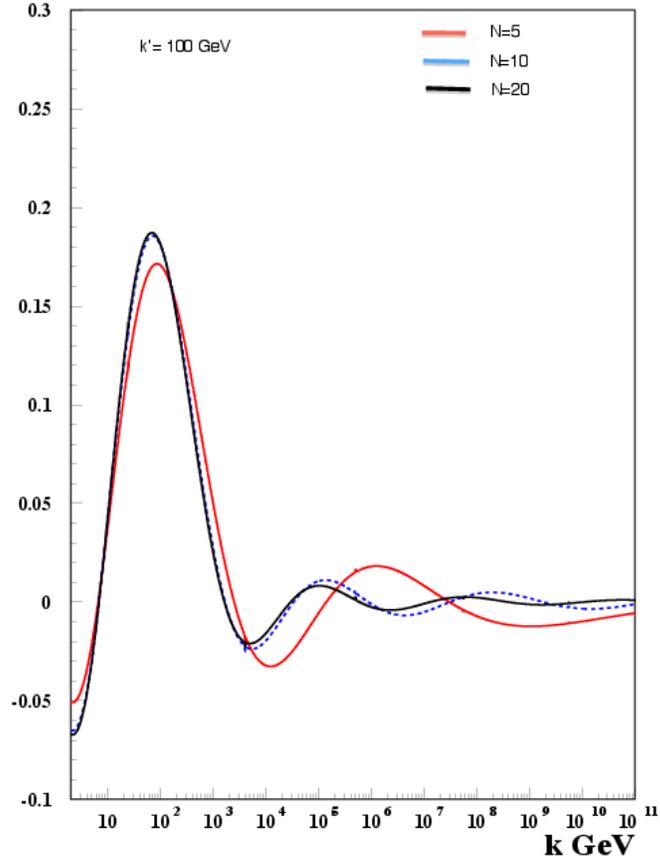} }
\caption{ The completeness sum shown as a function of $k_T$ and evaluated at $k_T'=100$ GeV, (\ref{complet}), for the first 5 eigenfunctions (red),
 first 10 eigenfunctions (blue) and first 20 eigenfunctions (black).}
\label{fig5} \end{figure}

For the completeness condition we consider the sum
\beq
 \sum_{n=1}^N f_{\omega_n}(t)f^*_{\omega_n}(t^\prime)  
 \label{complet}
\eeq 
For $k_T^\prime=100$ GeV we plot this sum in Fig.\ref{fig5}  for the 
$N \, = \, 5,10$ and $20$ eigenfunctions.

We see that the sum has converged\footnote{the fast convergence of this sum will be discussed below}
 after  10 - 20 eigenfunctions
to  a distribution on $k_T$ which is peaked at $k_T=k_T^\prime$.
However, we note that the distribution is very broad.
This tells us that the discrete eigenfunctions do {\it not} form a complete set by themselves. Rather
 the completeness requires  that the sum over the discrete eigenfunctions must be supplemented by the integral over
the continuum of states for which $\omega$ takes negative values.

For negative $\omega$ there is no critical transverse momentum, $t_c$, beyond which the eigenfunctions diminish,
but have  oscillatory behaviour for all $t$. We can write the negative omega eigenfunctions as
\beq f_{-|\omega|}(t) \ = \ \sqrt{\frac{2}{\pi}} \frac{1}{\sqrt{\alphabar(t) \chi^\prime\left(\nu_\omega(t)\right)}}
 \sin\left( \int_{t_0}^t \nu_\omega(t^\prime) dt^\prime + \eta_{NP}  \right). \label{negomega}  \eeq 
Since the ultraviolet boundary condition does not
impose a specific ultraviolet phase, the spectrum is continuous.\footnote{This is analogous to the fact that particle only form bound-states for negative energy.
 Here the analogue of energy is $-\omega$.}
 Although there is no ultraviolet phase-fixing condition, there can be an infrared boundary condition which 
determines the phase of the oscillations at small $t$. For small positive $\omega$ the
 eigenfunctions are very closely spaced 
 and become indistinguishable from a continuum. For small negative $\omega$, the non-perturbative phase should match
 its value for small positive $\omega$ in order to ensure a smooth function as $\omega$ changes sign.
 Note that  
for large negative $\omega$ it may be the case that the infrared phase is not defined. An example of a mechanism
in which this  happens is where the infrared behaviour of QCD is simulated by an effective gluon
mass \cite{LLS}. Here it is found that at some negative $\omega=-\omega_1$, there is a phase transition
 below which the infrared phase is not determined. Other possible sources of such phase transitions
 could arise from the restoration of conformal invariance at some high-energy scale.

\begin{figure}[htbp]
\centerline{\includegraphics[width=8cm]{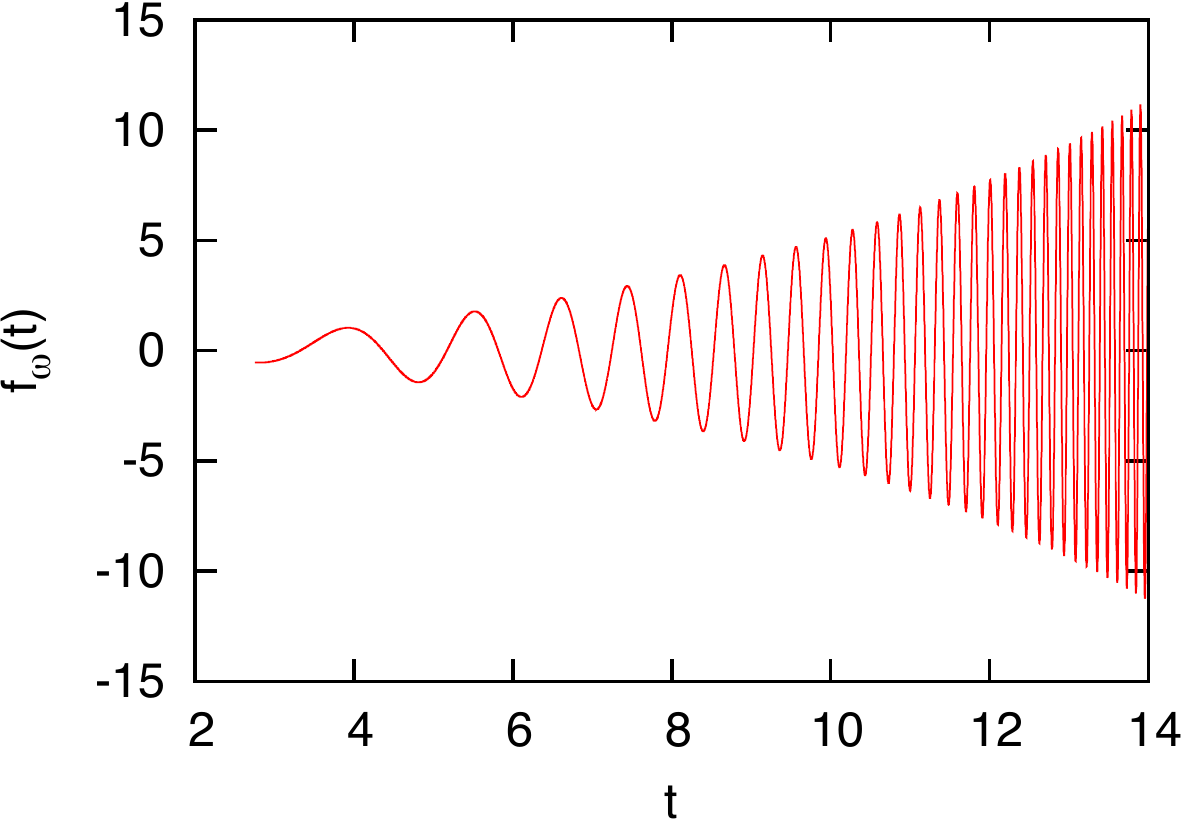} }
\caption{ The eigenfunction $f_\omega(t)$,  with $\omega=-1$, of eq.(\ref{negomega})  }
\label{fig5a} \end{figure}

For sufficiently large (negative) values of $\omega$, $\nu_\omega(t)$ 
  is given by
  \beq \nu_\omega(t) \ \approx
 \exp\left(-\gamma_E + \frac{|\omega|}{2\alphabar(t)} \right)   \eeq
and
 \beq \chi^\prime\left(\nu_\omega(t) \right) \ \approx \ 
  2\exp\left(\gamma_E - \frac{|\omega|}{2\alphabar(t)} \right),   \eeq     
from which we can see that for large $t$
these negative $\omega$ eigenfunctions have oscillations whose
 amplitude and frequency increase  rapidly as $t$ increases.
An example for which $\omega=-1$ is shown in Fig. \ref{fig5a}.

\begin{figure}[htbp]
\centerline{\includegraphics[width=9cm]{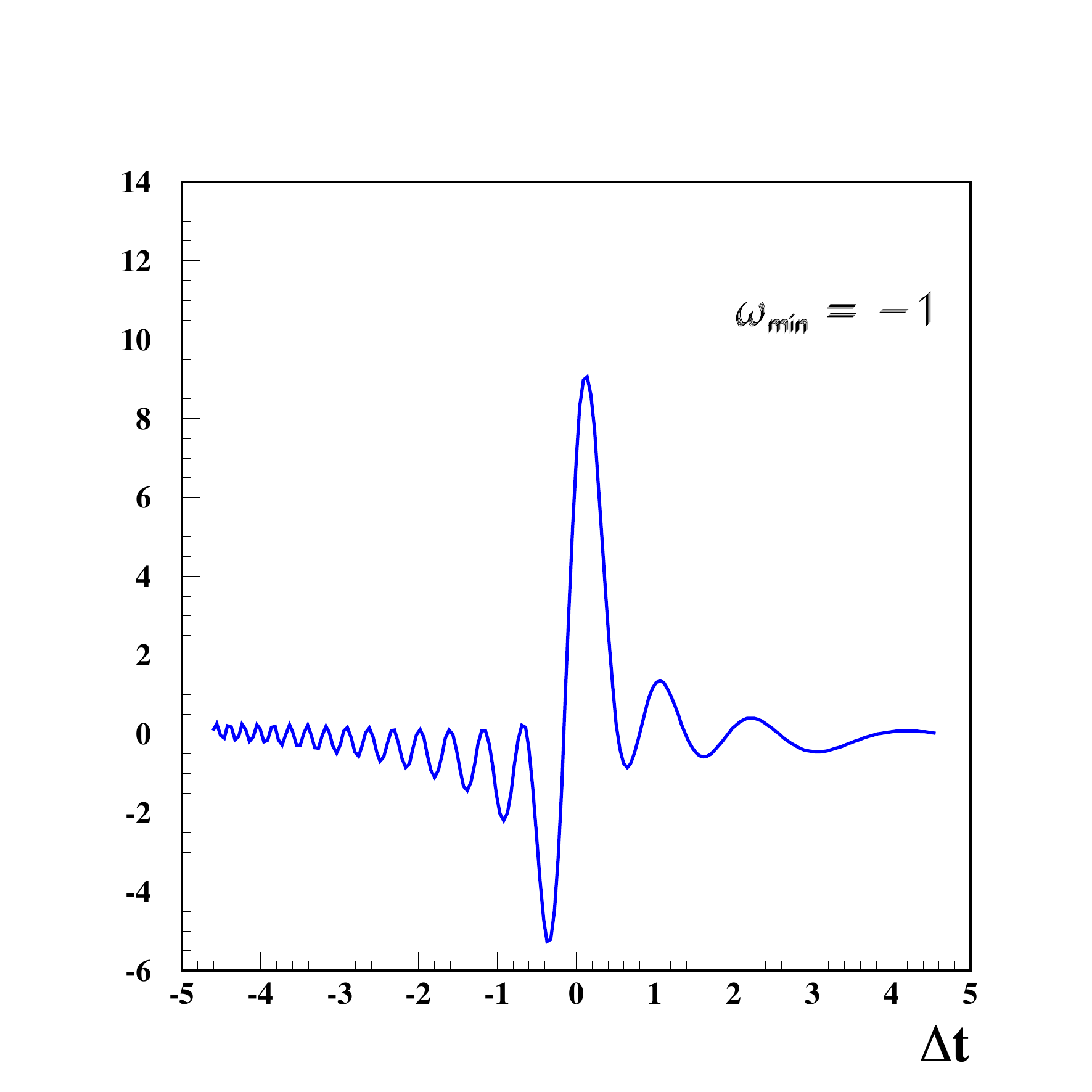} \
\includegraphics[width=9cm]{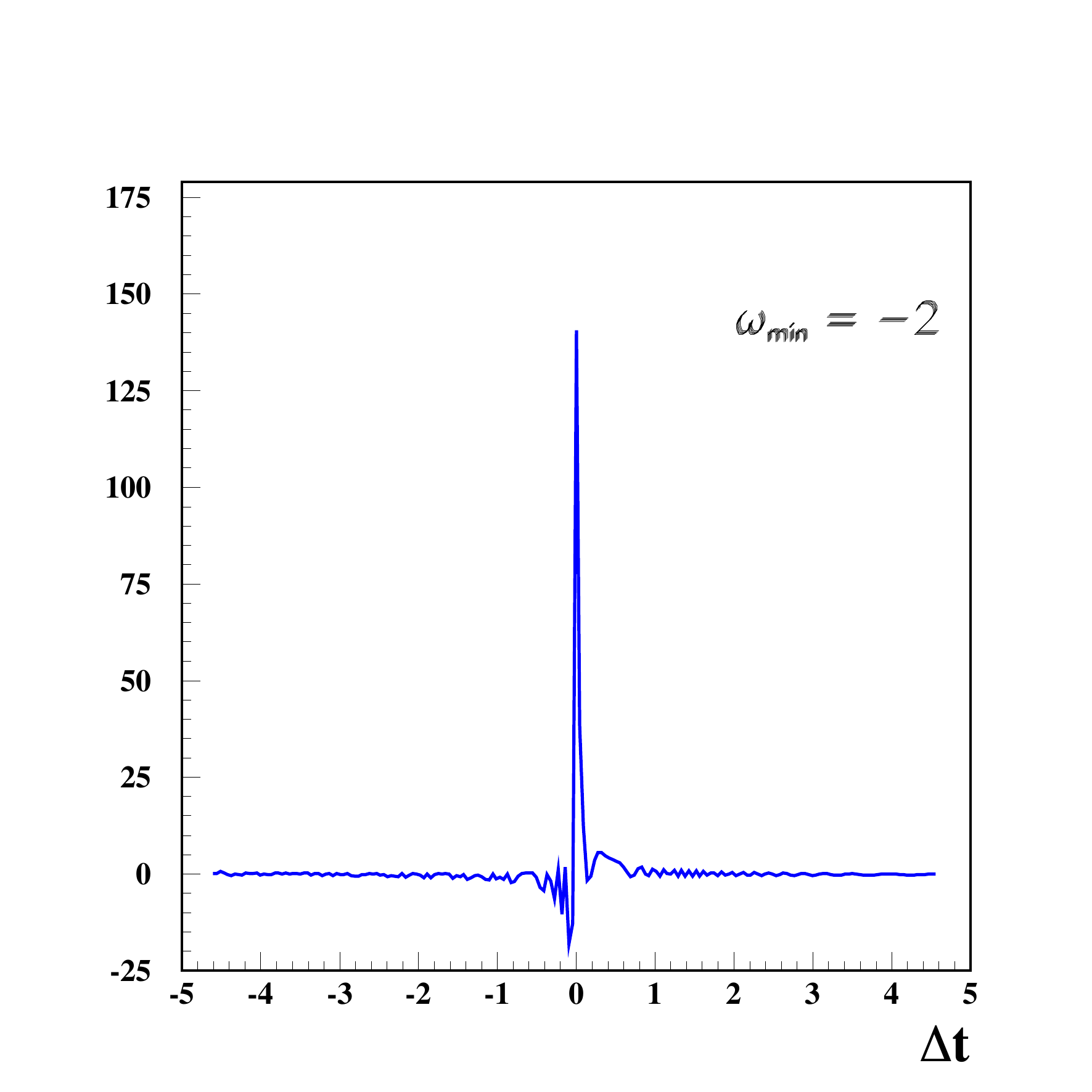}
 }
\caption{ The completeness relation (\ref{complete}) for $t'=10$,   with the negative $\omega$
 eigenfunctions included for  $\omega \, > -1$ (left)
 and   $\omega \, > -2$ (right).  }
\label{fig6} \end{figure}

With the inclusion of the continuum states, the completeness relation becomes
\beq \lim_{\omega_{min} \to -\infty} \int_{\omega_{min}}^0 d\omega f_{-|\omega|}(t) f_{-|\omega|}^*(t^\prime) 
 +  \sum_{n=1}^\infty f_{\omega_n}(t)f_{\omega_n}(t^\prime)   \ = \ \delta(t-t^\prime) \label{complete}  \eeq

In Fig. \ref{fig6} we plot this quantity for $\omega_{min}=-1$ and $\omega_{min}=-2$ for $t^\prime=10$.
We see that as $\omega_{min}$ becomes more negative the LHS of eq.(\ref{complete}) becomes more sharply
 peaked - tending to the required $\delta$-function in the asymptotic limit $\omega_{min} \, \to \, -\infty$.

\section{Transverse Momentum Dependence of the Residues of the Poles} \label{sec5}

As $t$ increases from $t=t_0$, (where the infrared phase is set), the eigenfunction $f_n$ oscillates
through $(n-1)$ nodes before the value of $t_c(n)$ for that eigenfunction,
with (in leading order) \beq t_c(n) \ = \ \frac{4\ln2}{\omega_n} \eeq 
whereas for  values of $t \, > \, t_c(n)$  it decays exponentially.

\begin{figure}[htbp]
\centerline{\includegraphics[width=8cm]{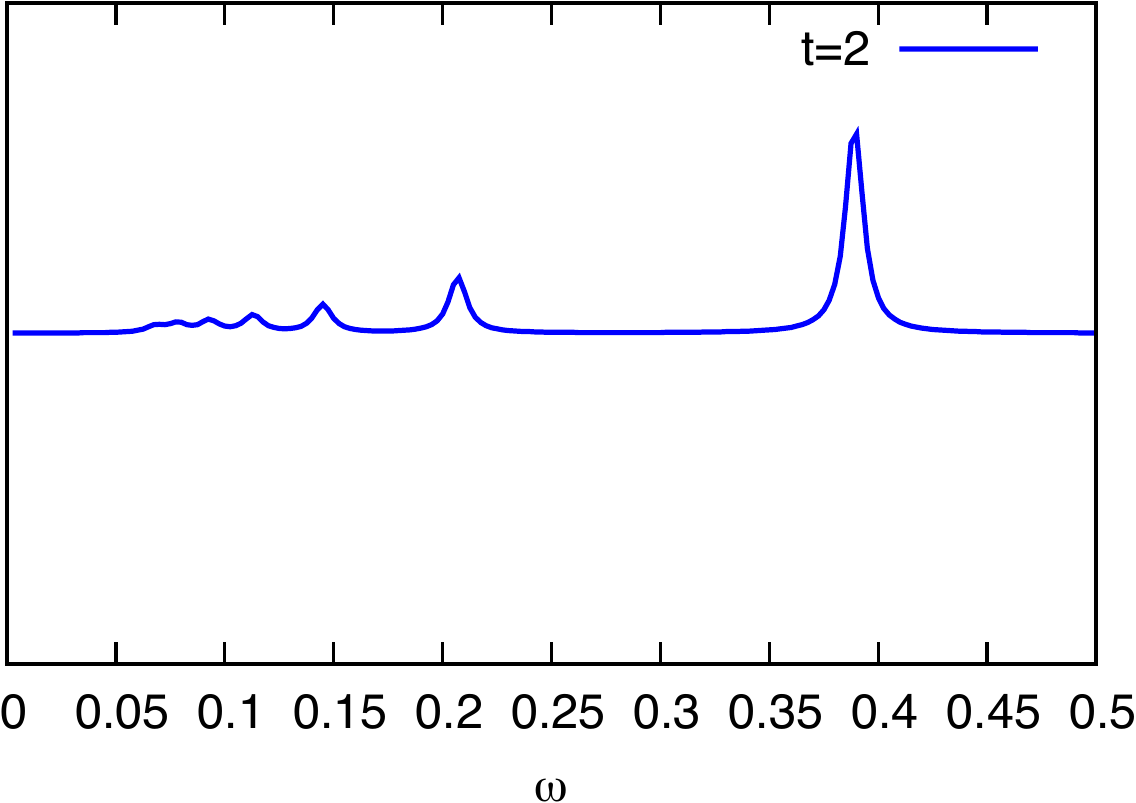} \
\includegraphics[width=8cm]{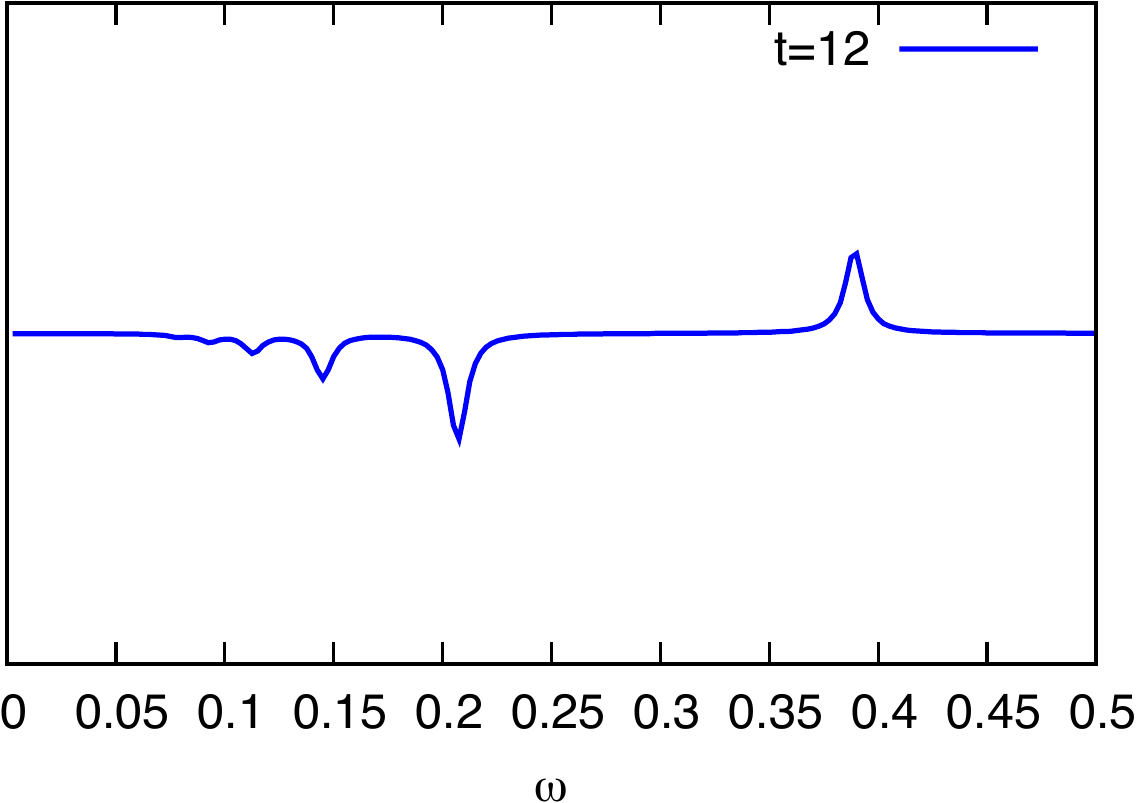}}
\caption{ The Green function  as a function of
 $\Re e \{ \omega \}$ with $\omega$ close to the real axis.
 Left panel $t=2,t^\prime=2$, right panel $t=12, t^\prime=2$ }
\label{fig7} \end{figure}
 
This means that for small $t$, $t \, \leq \,  t_c(1)$ 
 the Green function as a function of $\omega$  has a series of poles
 at $\omega =\omega_n$ with residues that oscillate with amplitudes that decrease with increasing $n$,
 reflecting a convergence of the sum over pole contributions.
 This is shown in the left-hand graph of Fig.\ref{fig7}, 
where we have taken $t^\prime=2$ (and $\omega$ taken close to the real axis).

In the region
$$ t_c(1) \ < \ t \ < \ t_c(2) ,  $$
the residue of the leading pole is attenuated and the leading behaviour is
 now given by the first sub-leading pole, which still has an oscillatory
 residue.  This is shown in the right-hand graph of Fig.\ref{fig7} for $t=12$.
We note that for this value of $t$ the residue of the leading pole
 has significantly diminished. Note also that the sign of the residue
 of the sub-leading pole is opposite to the case where $t=2$, reflecting
 the oscillatory nature of the residues of the poles for $t \, < \, t_c(n)$.

As $t$ increases further, the residues of more and more of the sub-leading
 poles start to decay and the inverse Mellin transform of the Green-function
 is dominated by the contribution from smaller and smaller values of $\omega_n$.

This is in sharp contrast to the situation in which the coupling is kept
fixed and for which the inverse Mellin
 (for large rapidity)  transform is {\it always} dominated
  by the region close to $4 \alphabar \ln(2)$. For the running coupling this
is not the case, but the value of $t$ at which the particular sub-leading poles
   dominates depend on the value of the rapidity, $Y$ (or $\ln(1/x)$
in the case of deep-inelastic scattering, leading to an effective
 pomeron intercept which depends on $Y$.

\section{Properties of the Unintegrated Gluon Density}
\label{sec6}
The Green function also has a spectral representation given  
by eq.(\ref{spectral}) and so we should be able to obtain a good approximation
 to the unintegrated gluon density by summing over the pole
 contributions from $n=1$ to $n=n_{max}$.

As a qualitative demonstration  of the unintegrated gluon density that
can be obtained from the BFKL Green function, we consider a very simple model
in which the non-perturbative phase, $\eta_{NP}$ is set to the constant value 
\footnote{This phase can, in principle,
 be $\omega$ dependent although it must lie within the range 
$-\frac{\pi}{2} < \eta_{NP} < \frac{\pi}{2}$ to avoid cross-over between
 eigenvalues. The freedom to select this phase as a function of $\omega$
 is likely to be necessary to get a fit between the BFKL formalism 
 and experimental data.} of $-\frac{\pi}{4}$. 
The proton  
impact factor has to be  positive everywhere and concentrated at the values of $k_T< {\cal O}(1)$ GeV.   It is usually assumed to  be of the form 
\beq \Phi_p(k_T) \ = \ A k_T^2 e^{-b k_T^2}, \label{protff}  \eeq 
 as discussed in ref. \cite{KLRW,KLR}. The form~(\ref{protff})  vanishes  as $k_T^2$ for small $k_T$,
 as required by colour transparency 
 and the coefficient $b$ has the interpretation of the average inverse
 square transverse momentum of partons inside the proton and is therefore of the order of 10 GeV$^{-2}$. The overlap integral of the proton impact factor with the eigenfunction for $t > t_0$ ($t_0$ corresponds to $k_T = 1$ GeV) is therefore determined by the value of the amplitude at $t=t_0$. This is due to the fact that the proton impact factor falls for $t \, > \, t_0$
at a rate which is much faster than the oscillation frequency of the eigenfunctions in the region $t \, \sim \, t_0$.
In ref  \cite{KLRW,KLR}  also  other forms of the proton  impact factor were investigated, e.g. with different powers of $k^2$ in the prefactor and/or the exponent.
It was  found, however,  that the fit to data has no sensitivity to such alternatives due, again, to a small range of the impact factor in comparison with the rate of change of the eigenfunction amplitudes. 
 Therefore,  for the purpose  of this paper  it is  sufficient to  take an impact factor which have support only at $t\, = \, t_0$.

%



\begin{figure}[htbp]
\centerline{\includegraphics[width=9cm]{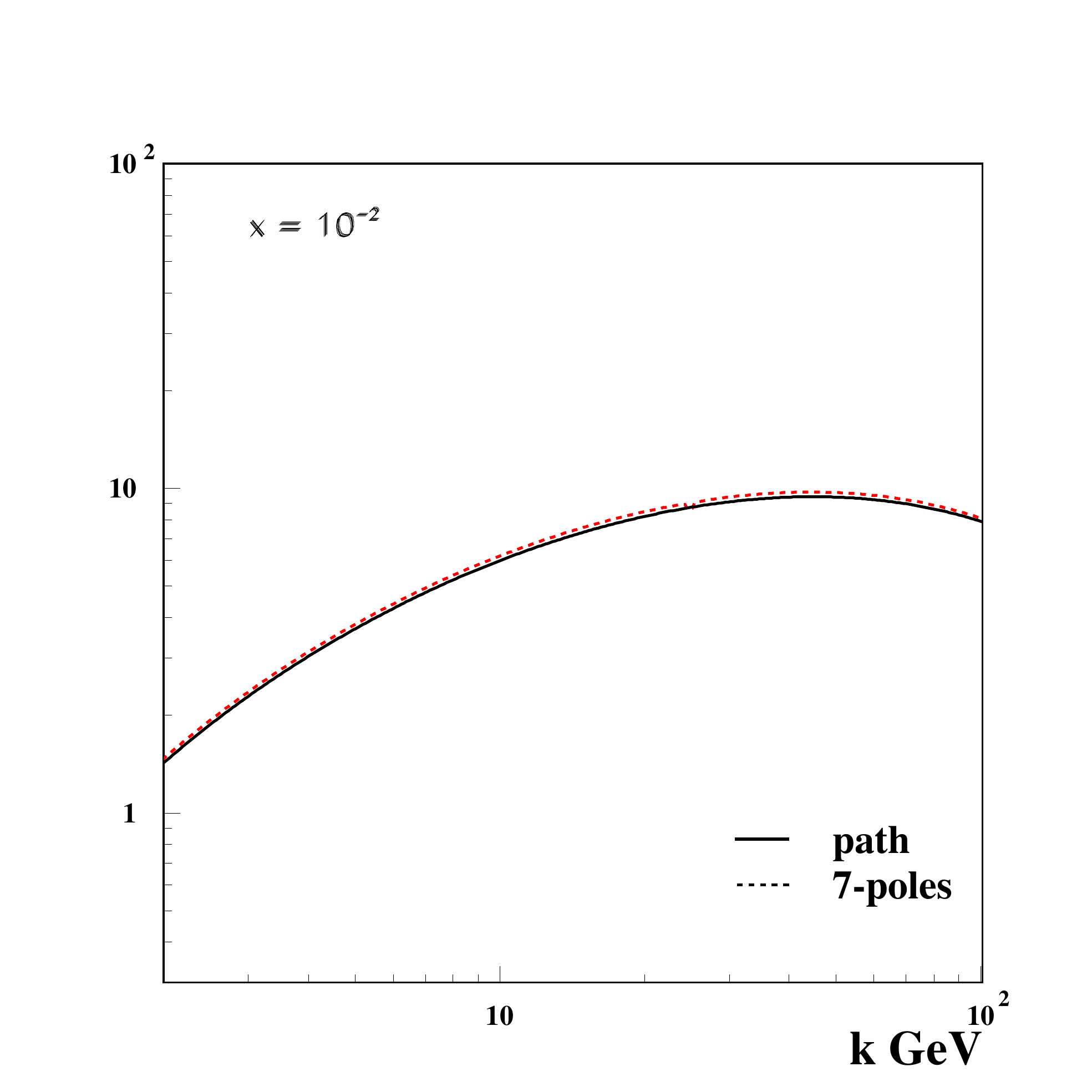}} 
\caption{ The unintegrated gluon density (\ref{gdot}), plotted as a function of $k_T$, for $x=10^{-2}$ and   
 $x=10^{-3}$ determined with the path (solid line) and pole (dashed line) evaluation. The pole result is increased by 3 percent for visibility.  }
\label{fig9} \end{figure}
With these parameters we plot the unintegrated gluon density
by performing the inverse Mellin transform given by eq.(\ref{gdot})
over one of the contours shown in Fig. \ref{twopaths}
 and compare it with the result obtained from the summation over the first 7 poles. We plot this in 
Fig.~\ref{fig9} for two different values of $x$, namely
 $x=10^{-2} $ and $  x=10^{-3}$. The pole result was increased by a factor of 1.03 for visibility. Without this increase both results would be indistinguishable. This perfect agreement was obtained because the path enclosed all the poles used in the sum. The integration over the path of Fig.~\ref{twopaths}   was performed down to $\omega_{min} =0.065$, which is between the $\omega_{min}$ values  of the 7th and 8th pole.    We checked this agreement for other values of  $\omega_{min}$ and the corresponding sum of poles and obtained  an equally good agreement.

This perfect agreement is, of course, due to Cauchy's theorem. This agreement is however non-trivial because the path integration is not closed what means that the missing piece of the path gives a negligible contribution, which is a first sign of a very good convergence of both the path and pole computations. In addition we note that both computations are numerically very different so it is therefore a very good check of the computational accuracy.   
 
 Another reason for this very good agreement is  the fact that we limited the comparison to the region of relatively small virtualities, $k_T<100$ GeV.  At $x=10^{-2}$,
 this region is experimentally relatively well accessible  at LHC and possible future colliders. At $x=10^{-3}$ and for smaller $x$ values the experimentally accessible $k_T$ region diminishes substantially.
  Notwithstanding this, the subleading poles turn out to have a measurable effect (as we shall discuss
 in subsection \ref{sec7.4}) and this will be significant for the prospect,
 discussed in section \ref{sec9}
 of the identification of new physics from the shift in positions and residues of sub-leading poles.

\begin{figure}[htbp]
\centerline{\includegraphics[width=9cm]{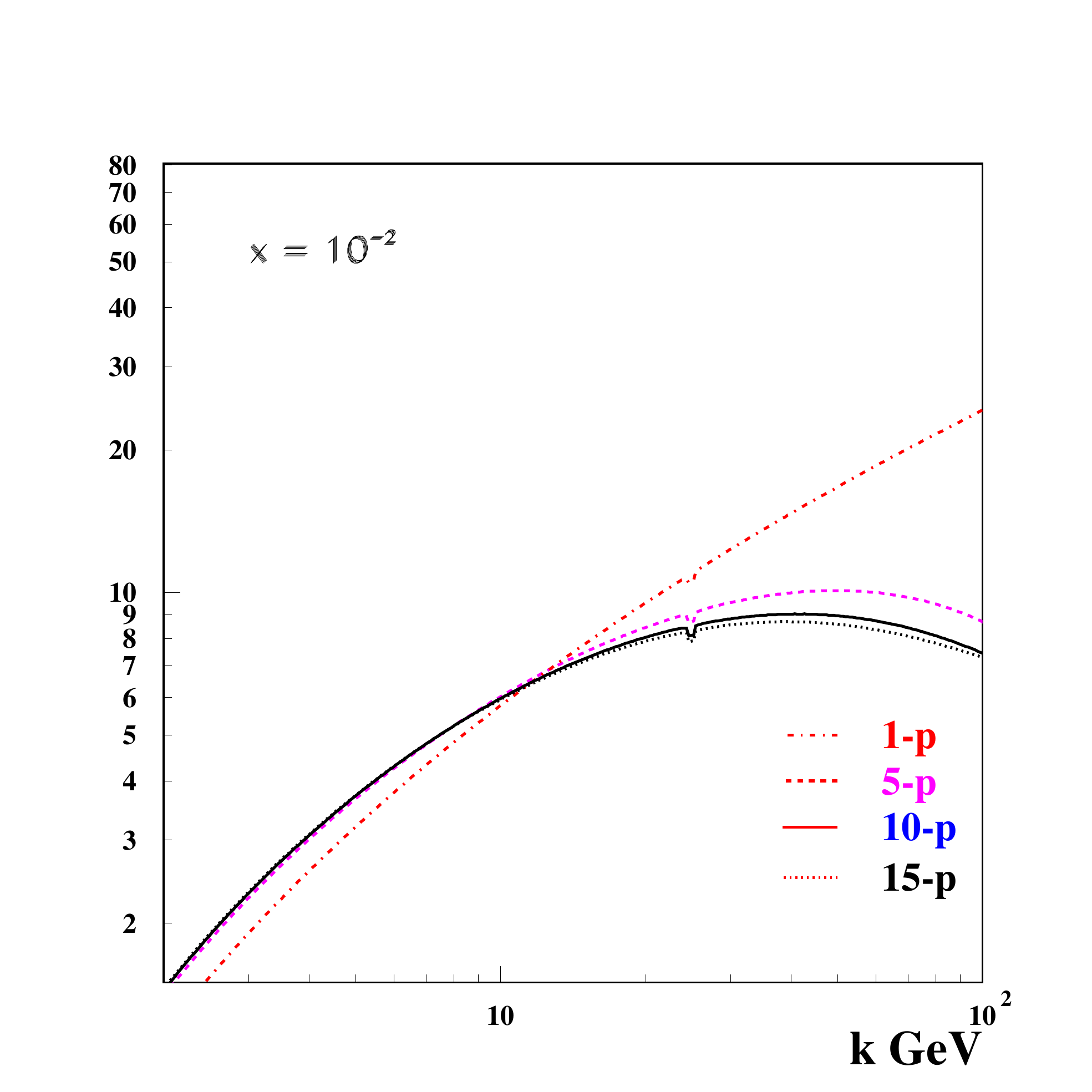} 
\includegraphics[width=9cm]{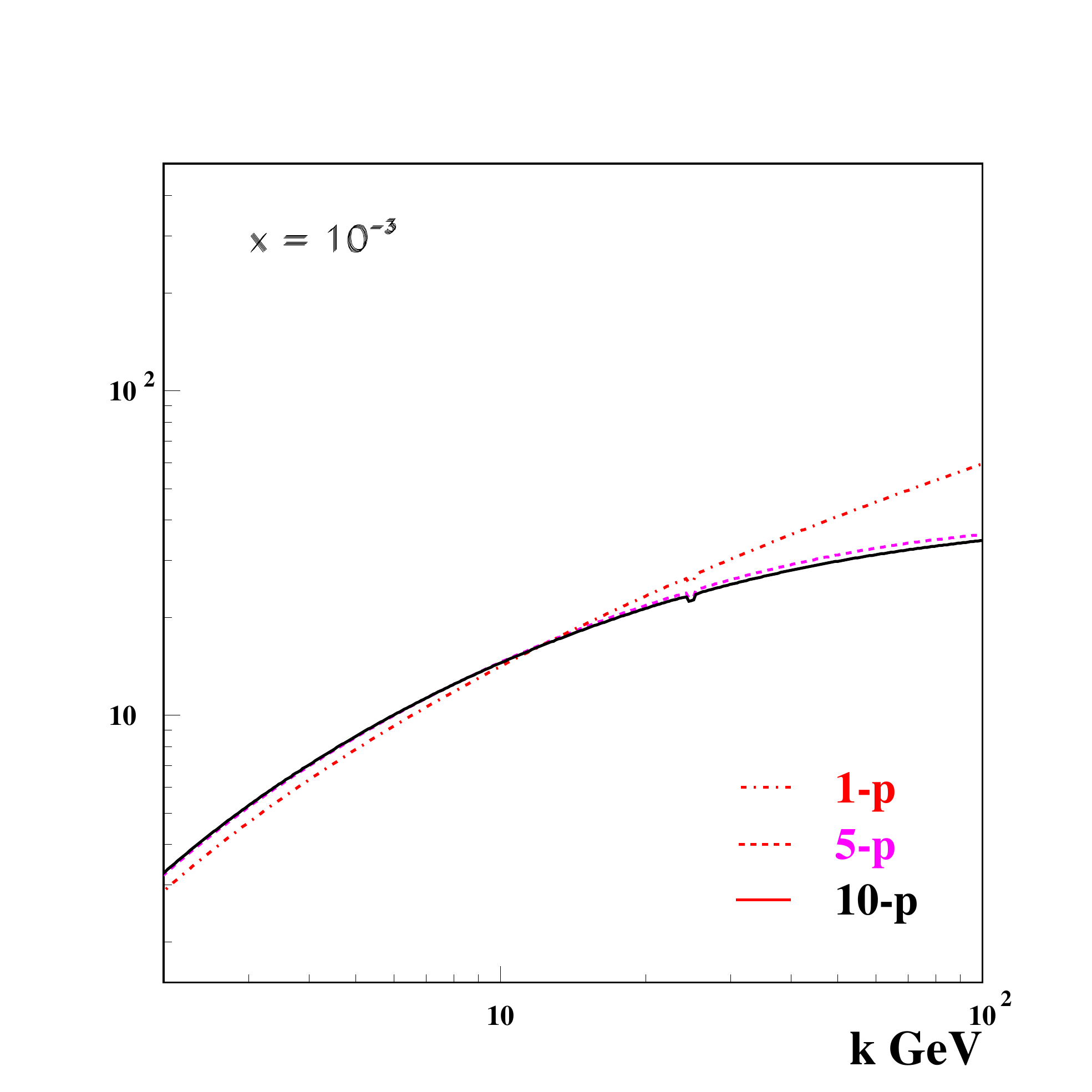}}
\centerline{\includegraphics[width=9cm]{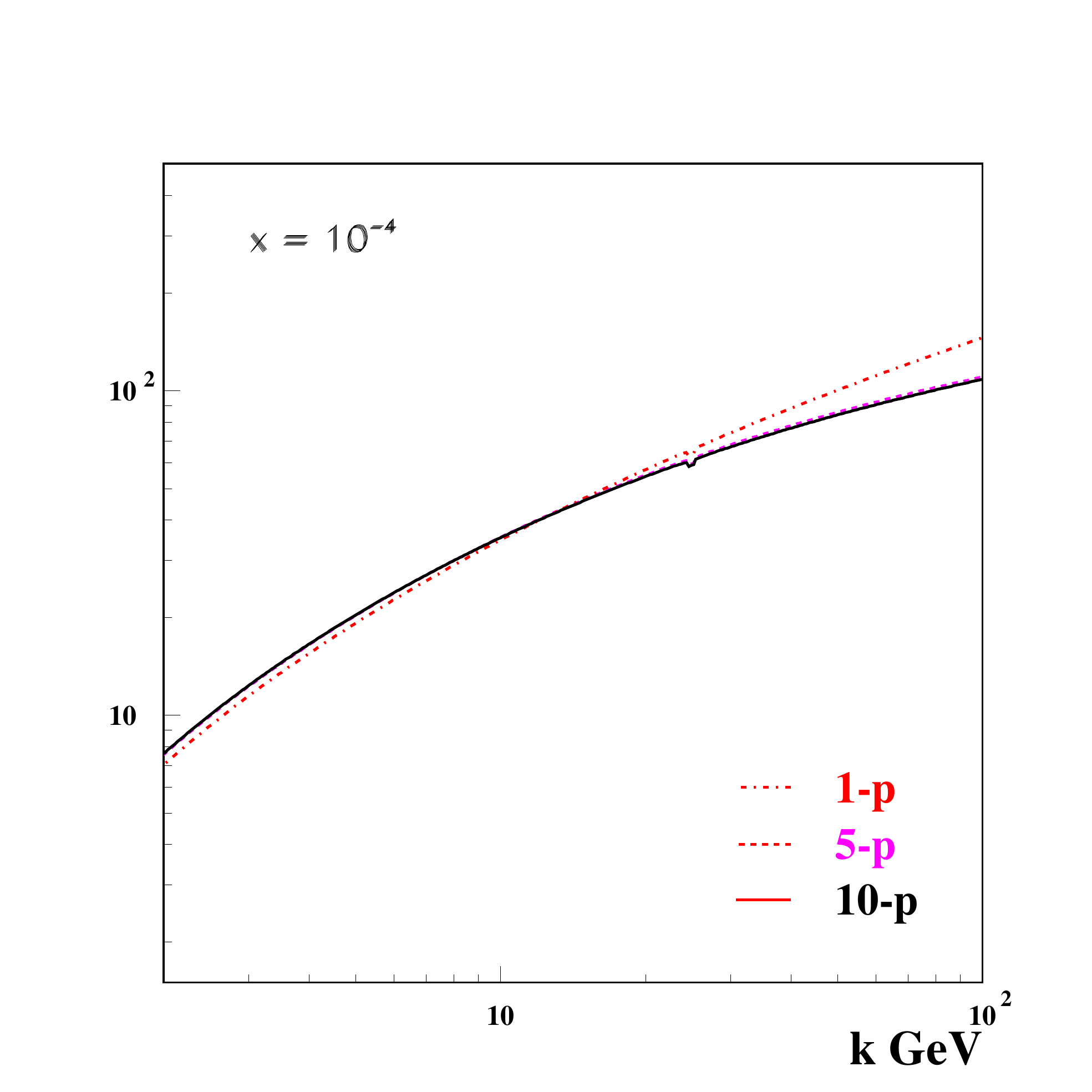}}  
\caption{ The unintegrated gluon density (\ref{gdot}), plotted as a function of $k_T$,  for $x=10^{-2}$ and   
 $x=10^{-3}$ determined with the pole method. The dashed-dotted line shows the leading pole contribution, the dashed line shows the sum of 5 poles, the solid line the sum of 10 poles and the dotted line the sum over 15 poles.}
\label{figpoles} \end{figure}

 In the low $k_T$ region, as first indicated in Fig.~\ref{fig9}, we have a fast convergence of the gluon density as a function of the number of poles used in the summation.
 In Fig.~\ref{figpoles} we show the gluon density computed with different number of poles. The dashed-dotted line shows the leading pole contribution, the dashed line shows the sum of 5 poles, the solid line the sum of 10 poles and the dotted line the sum over 15 poles.  We see that $x=10^{-2}$ the first 10 poles almost converges in the whole $k_T$ region, for $x=10^{-3}$ the 10 and 15 pole summation are
 completely indistinguishable.  
We note also that for transverse momenta below around 10 GeV for the case $x=10^{-2}$
 and around 20 GeV for smaller values of $x$, the leading pole provides a reasonable approximation.

\subsection{Convergence of the Sum Over Poles}
\label{sec7}

We can understand how the pole sum converges by considering the behaviour
 of the normalized eigenfunctions for large $n$. In this region
 the eigenvalues are very small ($\omega_n \, \sim  \, 1/n$)
  and the critical momentum $t_c$ is very large (proportional to $n$).
 This means that for accessible values of $t$ the 
  RHS of eq.(\ref{freq}) is very small and eigenfunctions
 oscillate with approximately  a fixed frequency, $\nu_0$, given by
  $$\chi(\nu_0) \ = \ 0. $$
In this region we may use eq.(\ref{greenosc}) 
and the fact that the phase of the oscillation is $\eta_{NP}$ 
 at $t=t_0$ to write $\phi(\omega)$
 as
 \beq \phi(\omega) \ \approx \ \eta_{NP}-\frac{\pi}{4}-\nu_0t_0
 + \frac{1}{\betabar \omega} \int_0^{\nu_0} \chi(\nu^\prime) d\nu^\prime 
 \eeq
so that
 \beq \phi^\prime(\omega) \ \sim \frac{1}{\omega^2} \ \sim \ n^2 \eeq
The normalization factor $N_\omega(t)$ given by eq.(\ref{norm1})
 has a numerator factor $|z(t)|^{1/4}$ which cancels an identical
 factor in the denominator of the Airy function $Ai$  for $t \, \ll \, t_c$
 and the remaining factor is approximately $n$ independent as we replace
  $\nu_\omega(t)$ by $\nu_0$ in the argument of $\chi^\prime$.

The upshot of this is that the two factors of $1/\sqrt{\phi^\prime(\omega)}$
(see eq.(\ref{normeigen})) give rise to a convergence 
of the eigenfunction series at  small $t$ (relative to $t_c$) like  $\sim 1/n^2$.  Since $t_c$ increases quickly with the eigenfunction number $n$ this fast convergence always happens in the experimentally accessible region of $t$.

\subsection{The Role of the Continuum for Negative $\omega$}
\label{sec8}
We have seen in section \ref{sec4} that the contributions from eigenfunctions
 with negative eigenvalues (i.e. negative $\omega$) are essential in order to
 provide a complete set of eigenfunctions obeying the closure relation
(\ref{complete}). In this section we discuss the effect that the Green function
with
negative $\omega$  has on the unintegrated gluon density.

The contribution to this expression for the  unintegrated  gluon density
 from the (positive $\omega$)   poles vanishes asymptotically with $t$,
  reflecting the behaviour described in section \ref{sec5} whereby as $t$ 
increases the residues of more and more of the poles pass from 
 an oscillatory behaviour to an exponentially damped behaviour.

The contribution, $\Delta_{\omega<0} $ to $\overline{g}(x,k)$
from negative $\omega$ is given by
\beq 
\Delta_{\omega<0}  \overline{g}(x,k) \ = \
 \lim_{\omega_{min}\to -\infty} \int_{\omega_{min}}^0
  d\omega x^{-\omega} \int dt^\prime
     f_{-|\omega|}(t) f_{-|\omega|}(t^\prime) 
   \Phi_P(t^\prime),  \label{gbarnegomega} \eeq
with $f_{-|\omega|}(t)$ given by eq.(\ref{negomega}) (and $t=2\ln(k_T/\Lambda)$).

\begin{figure}[h]
\centerline{\includegraphics[width=8cm]{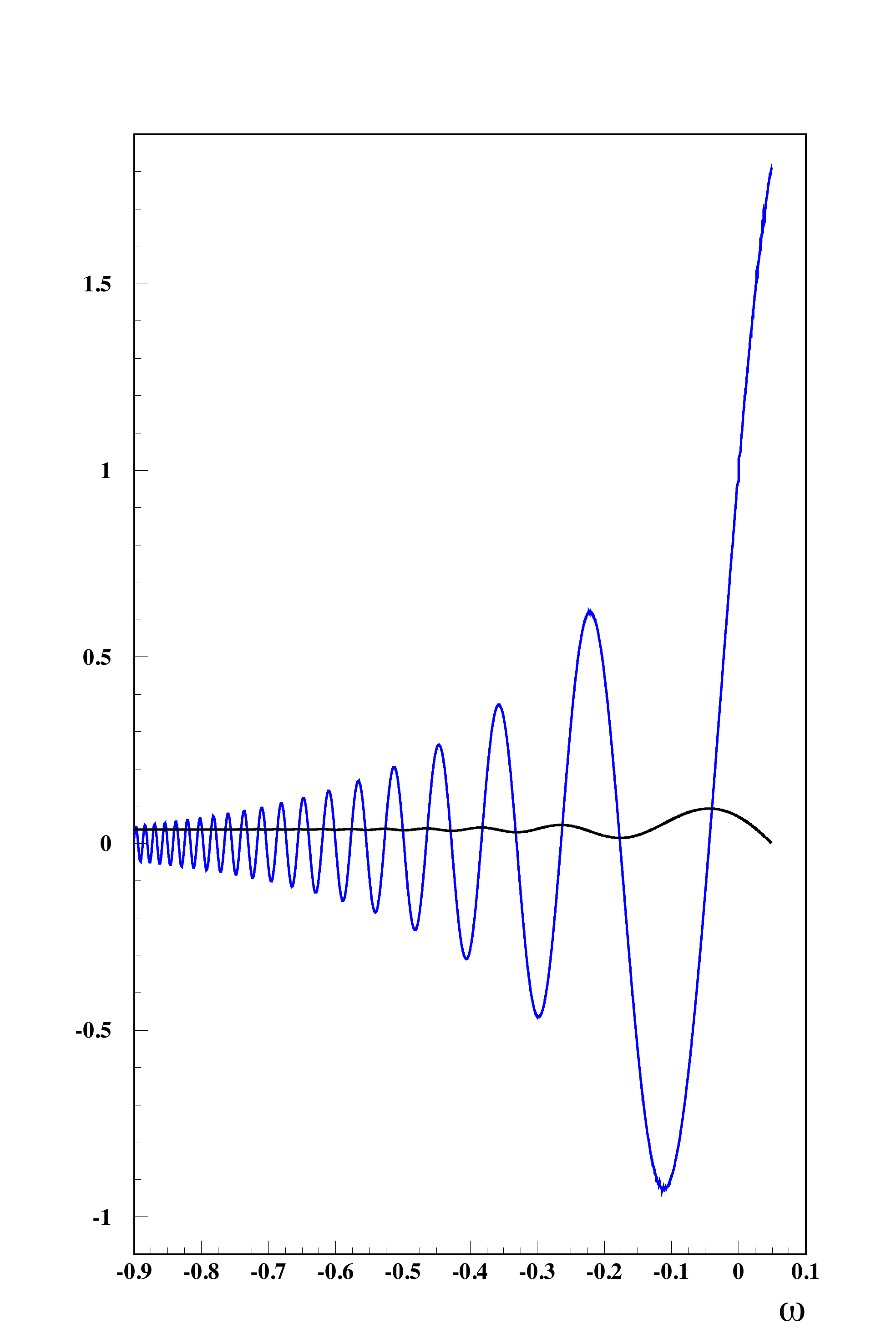}}
\caption{Blue line: The integrand for the RHS of eq.(\ref{gbarnegomega}).
    Black Line: The integral from $\omega=0$ to  $\omega_{min}=\omega$.
We see that the integral has converged well for $\omega_{min}=-1$.
We have taken $x=10^{-3}$ and $t=12$ }
\label{negeigf} \end{figure}

 At first sight, one would expect this contribution to have a negligible 
effect on the unintegrated gluon density owing to the factor of $x^{\omega}$.
However, as can be seen from Fig. \ref{negeigf}, for values of $\omega$
 just below $\omega=0$ the integrand of the RHS of eq.(\ref{gbarnegomega})
 is still quite large and rapidly oscillating, although we can also see that
the integral converges to a fairly small value by $\omega=-1$.


\begin{figure}[htbp]
\centerline{\includegraphics[width=9cm]{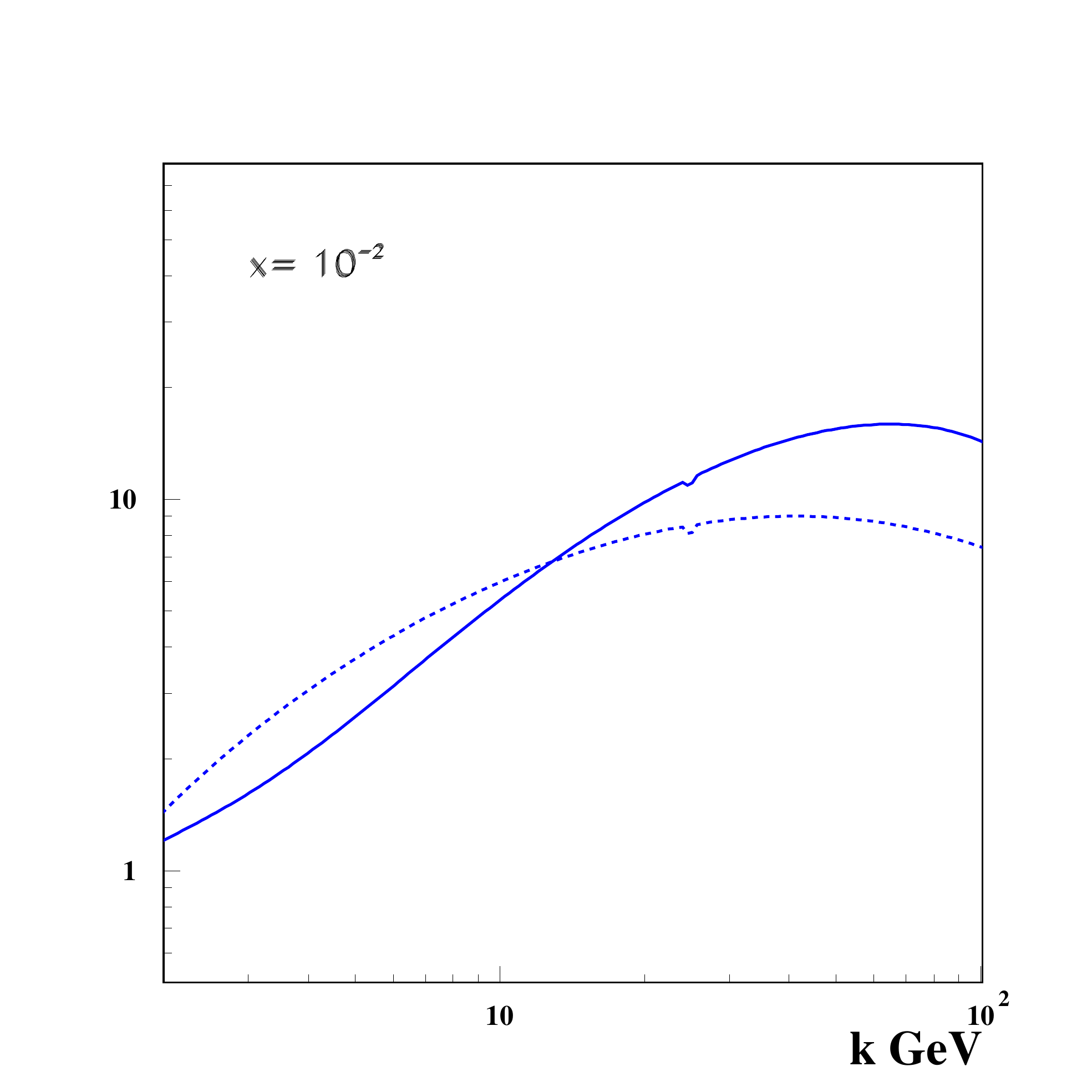} 
\includegraphics[width=9cm]{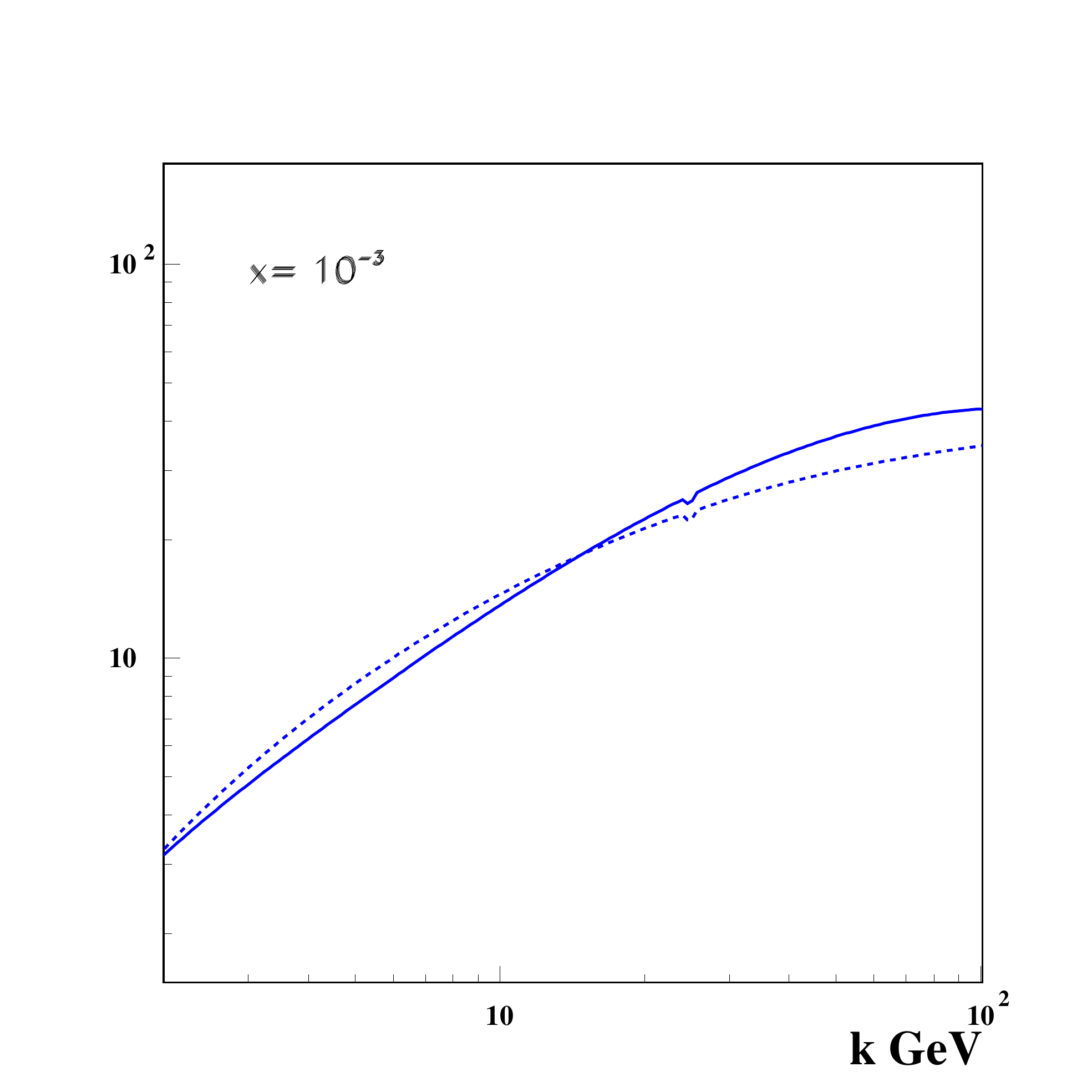}}
\centerline{\includegraphics[width=9cm]{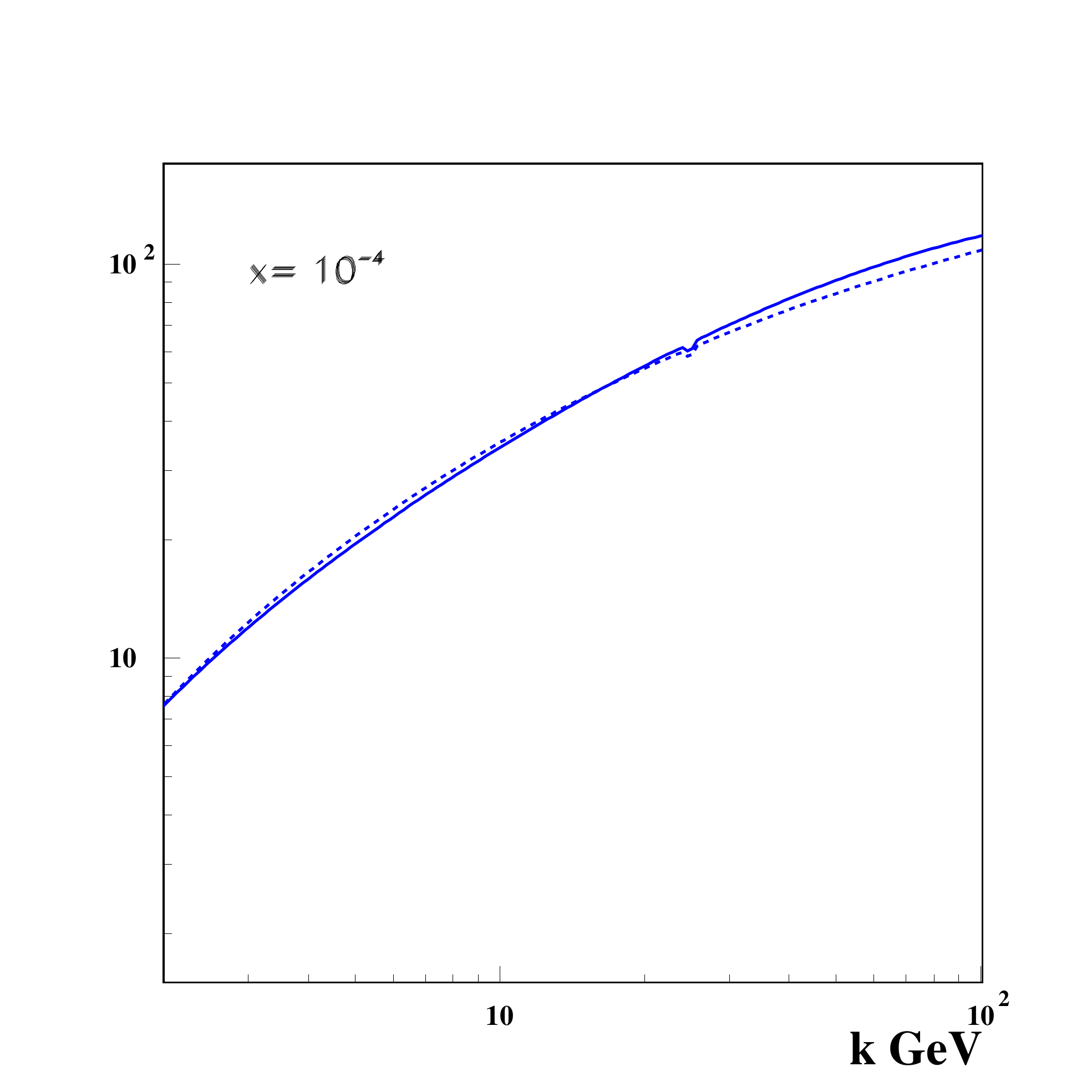}}  
\caption{ The unintegrated gluon density for $x=10^{-2}$,   $x=10^{-3}$ and
 $x=10^{-4}$ including  the contribution of negative $\omega$'s. The dashed line shows the pole contribution (computed from the sum of 10 poles), the solid line shows the same pole gluon density with added contribution of negative $\omega$'s. }
\label{figpolneg} \end{figure}

We have computed the gluon density including the  contribution from negative $\omega$. The contribution of positive  $\omega$'s was given by the sum of the first 10 poles and the contribution  between  $\omega=0$ and  $\omega_{10}$ (where the poles are densely packed) was treated as a continuum. The negative omega contribution was evaluated  from $\omega=-2$ to  $\omega=0$ using  eq. (\ref{gbarnegomega}). Fig. \ref{figpolneg} shows the unintegrated gluon density for $x=10^{-2}$,   $x=10^{-3}$ and
 $x=10^{-4}$ including  the contribution from negative $\omega$. The dashed line shows the pole contribution (computed from the sum of 10 poles), the solid line shows the same pole gluon density with added contribution of negative $\omega$'s. The negative $\omega$ contribution was computed assuming that the infrared  phase 
  for positive and negative $\omega$ match at $\omega=0$.

Fig. \ref{figpolneg} shows that the contribution of negative $\omega$ may be substantial at  $x=10^{-2}$ if the negative $\omega$ 
infrared phase is fixed and substantially different from 0.  
However, this contribution diminishes very fast with decreasing $x$ as can be seen in the same figure.

\subsection{Comparison with DGLAP}

We see from Fig.\ref{fig9} that for $x=10^{-2}$ above $k_T$ of around 30 GeV, the unintegrated gluon density
ceases to rise, whereas for smaller values of $x$, e.g. $x=10^{-3}$, the unintegrated gluon density
 continues to rise up to transverse momenta of 100 GeV. This can be understood from the $t$-dependence
of the residues of the poles as seen for example in Fig.\ref{fig7}. For $t=12$ (corresponding to $k_T$
of around  100 GeV), we see that the residue of the sub-leading pole is of opposite sign and
slightly  larger in magnitude that that of the leading pole (whose residue is evanescent since $t$
 is above $t_c$ for that pole). The contribution of the sub-leading pole is suppressed by a factor of
 $$ x^{(\omega_1-\omega_2)} \ \approx \ x^{0.2}. $$
For $x=10^{-2}$ the contribution from the sub-leading pole is sufficient at such values of $t$ 
to halt the rise in the unintegrated gluon density, whereas for $x=10^{-3}$ it is insufficient.
Nevertheless, at sufficiently large $t$ the unintegrated gluon density for $x=10^{-3}$ 
will also cease to rise and will eventually display  oscillations. 
Above some large $t$, outside  the  range of $t$ ( $k_T$) range considered in this paper, these oscillation are certainly unphysical because they lead to negative gluon density. 

There is no reason {\it a priori} why the BFKL amplitude should not display
oscillations. The inversion of the Mellin transform consists of an integral
over $\omega$ which has greatest support at the saddle-point $\omega_s$.
For values of $t$ below $t_c$ for this value of $\omega$ the amplitude
displays oscillatory behaviour and it is only when $t$ exceeds this critical
value that
the oscillations halt.  
 The unphysical oscillatory behaviour
 indicates that the solution to the BFKL equation is being 
applied to deep-inelastic scattering outside the kinematic range for which
it was intended. The application of the
 gluon  scattering amplitude in the Regge  regime to the 
 determination of the unintegrated gluon density  identifies the rapidity
 with  $\ln(1/x)$, which in LO is only valid
 provided the rapidity   significantly  exceeds $(t-t^\prime)$. 
 Therefore for $x=10^{-2}$ and $t^\prime$ confined to the low $t$
region where the proton impact factor has non-negligible support, we would expect the BFKL determination
of the unintegrated gluon density to become invalid if $t$  is substantially larger than $\sim \,5$ (corresponding
 to $k_T$ of order 10 GeV).

 We would expect this limitation on the allowed range of $t$ to be less stringent when NLO effects are taken into
account. Indeed, as pointed out 
 by Salam \cite{salam}, the LO treatment ignores the discrepancy between the rapidity variable used in a BFKL analysis
 (which is symmetric in $t$ and $t^\prime$) and the variable $x$. This introduces a factor of
 $$ e^{\omega(t-t^\prime)/2}, $$
whose absorption generates the largest part of the NLO contribution to the  characteristic function, $\chi(\nu)$.
We would therefore expect the BFKL amplitude computed at NLO to be less sensitive to the difference $(t-t^\prime)$
than a purely LO analysis.

It is known that at sufficiently large $t$ and sufficient small $x$, the double
 logarithm limit (DLL) of a BFKL analysis matches that of a DGLAP analysis.
  In Mellin space, the region in which this approximation is valid is
given by
   $$ 1 \ \gg \ \omega \ \gg \ \alphabar(t).  $$
In terms of $x$ this translates into the limits
 $$ \alphabar(t) \ln\left(\frac{1}{x} \right) \ \ll \ 1, \ \ \ 
 \ln\left(\frac{1}{x}\right)  \ \gg \ 1   $$
Moreover the match between a DGLAP analysis and a BFKL analysis
 can only be valid if $t$ exceeds $t_c$ at $\omega \approx \omega_s$,
where $\omega_s$ is the saddle-point for the inversion of the Mellin transform,
 i.e. the region around $\omega_s$ is where the integrand has its maximum
 support (assuming that this saddle-point lies to the right of all poles).
We have seen that we need to have values of $x$ smaller than $x=10^{-3}$
 in order to avoid a signal from the oscillatory part of the
 BFKL eigenfunctions. This means that 
 in the case of a DGLAP analysis, the DLL limit can only be reached
 for very small values of $\alphabar(t)$, i.e. values of $t$ way beyond
 any reasonable experimentally accessible region. 
 For the case of BFKL
 with running coupling  we need to go to even smaller values of $x$
 before the DLL becomes a reasonable approximation. This is because for 
running coupling the contributions from leading poles are attenuated
 at large $t$ and we need to be at sufficently large rapidity
 to ensure that these leading poles dominate the unintegrated 
  gluon density.  The input for a DGLAP fit is the structure function at some reference
 photon invariant mass, $Q_0$. In the case of the discrete BFKL pomeron the input 
 would be the proton impact factor and the infrared phase $\eta_{NP}$. As discussed in section \ref{sec6}
 this impact factor is expected to be dominated by its value at $t=t_0$ so that the only free parameter associated
with this impact factor would be the overall normalization. 
The only other parameter which can be substantially varied is the  infrared phase, $\eta_{NP}$ ,which should be a function of the eigenvalue $\omega_n$. The infrared phases  $\eta_{n}$  are  generated, as the eigenvalues  $\omega_n$,  by the quasi-bound states of gluons inside the proton and therefore  should be
 described by a simple parameterization. Our previous experience indicate that  two parameters may be sufficient in order to generate the $\eta - \omega$ dependence  required to fit data.

Because of these very different parameterizations, it is quite likely that there exists an overlap kinematic region
at low-$x$ for which data can be equally well described by a conventional DGLAP fit or a fit to the discrete BFKL  pomeron. A detailed comparison of the 
discrete BFKL pomeron with data will be performed in the next paper, after the NLO corrections are introduced. In the same time we will discuss the comparison with the DGLAP fit.   
 


\subsection{Dependence on  $x$ }
\label{sec7.4}

\begin{figure}[h]
\centerline{\includegraphics[width=9cm]{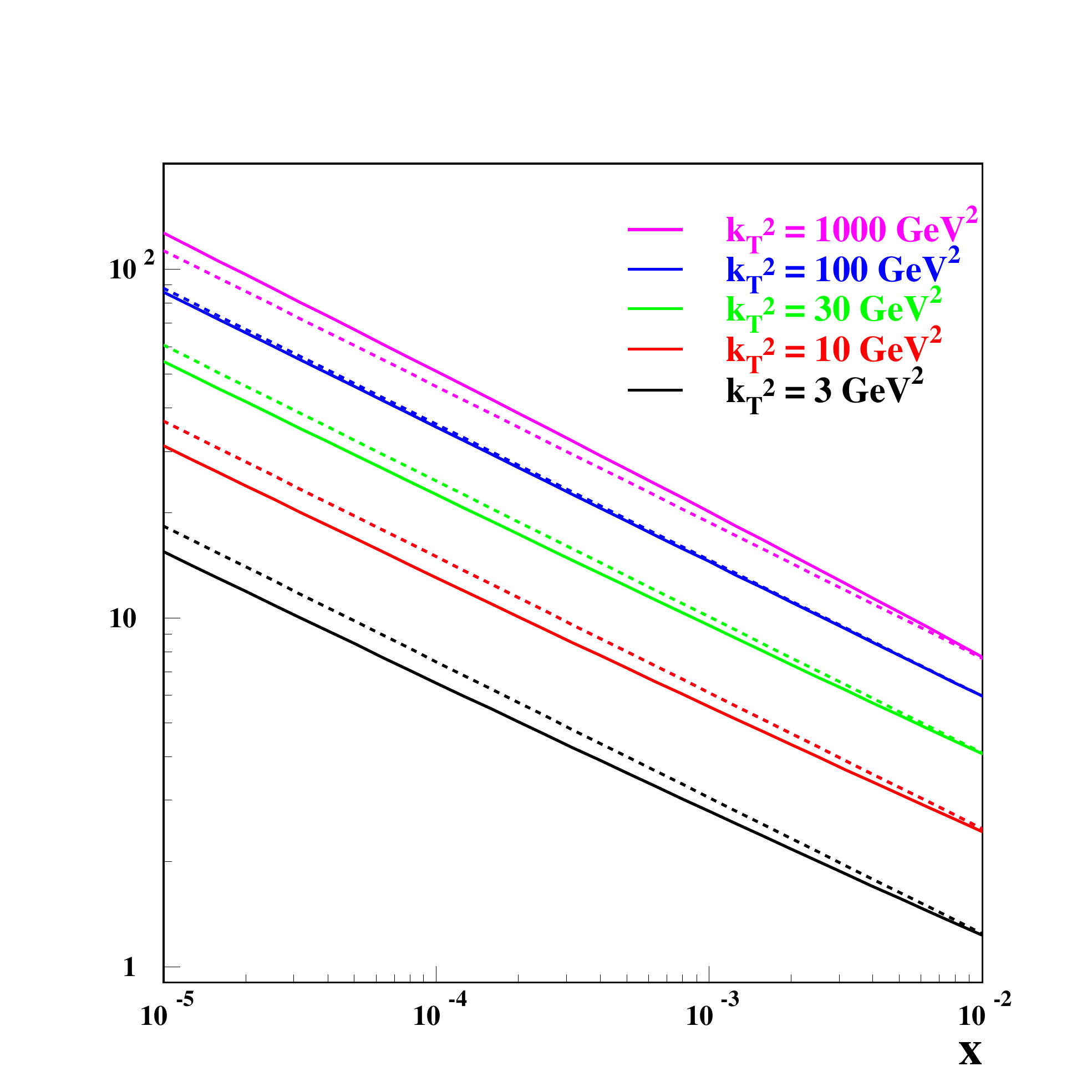}} 
\caption{ The unintegrated gluon density as a function of  $x$  determined from the pole contribution only, at various $k_T^2$. The dashed line shows the leading pole contribution normalized to the values of the gluon density at $x^{-2}$, for each $k_T^2$.}
\label{figx-poles} \end{figure}

\begin{figure}[h]
\centerline{\includegraphics[width=9cm]{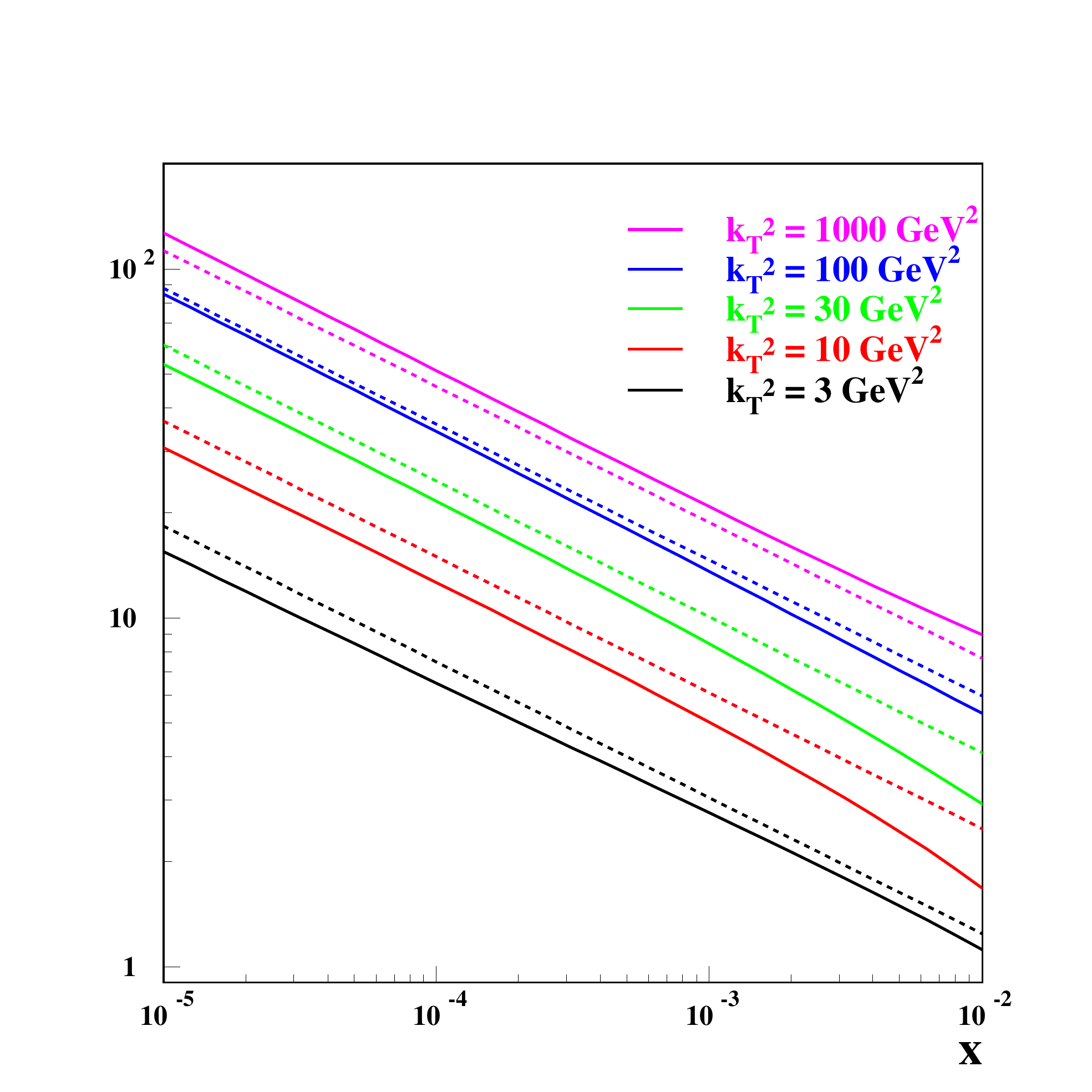}} 
\caption{ The unintegrated gluon density as a function of  $x$  determined from the pole contribution and the negative $\omega$ contribution, at various $k_T^2$. The dashed line shows the leading pole contribution, normalized in the same way as in Fig. \ref{figx-poles}.}
\label{fignx-poles} \end{figure}

It is  well known from HERA data that the $x$ dependence of $F_2$, which is directly  connected to the gluon density, exhibits a striking Regge type behaviour, i.e. $\sim (1/x)^{\lambda}$. The parameter $\lambda$ is not a constant, it increases logarithmically with $Q^2$ (which we set here equal to $k_T^2$), see e.g. \cite{Cald}.  
Such a behaviour is also a feature of the gluon density obtained from the Green function solution of BFKL  investigated here, see Fig. \ref{figx-poles}.  
The figure shows the unintegrated gluon density as a function of  $x$,  determined from the pole contribution only, for  various values of $k_T^2$. 
In this log log plot the function $(1/x)^{\lambda}$ is a straight  line, so we see immediately that the gluon densities  exhibit the same striking linearity as data.
The slope $\lambda$ increases slightly with $k_T^2$ which can be seen by comparison of the slope of  gluon density (full line) with the  $k_T^2$-independent slope of the  
 leading pole contribution (dashed line). The leading pole contribution is,  for each $k_T^2$, normalized to the values of the gluon density at $x=10^{-2}$.

In Fig. \ref{figx-poles} we display the gluon density in the validity region  of our solution to BFKL, $\Delta t < \log(1/x)$, which means that $k_T$ should be smaller than  order of 10 GeV at $x = 10^{-2}$. At larger $k_T$, not shown in this figure, we  observe a clear deviation from linearity in the region of $x$  between $10^{-2}$ and $10^{-3}$. A close inspection of the highest $k_T^2$ line in Fig. \ref{figx-poles} shows that  this effect sets in already at a relatively low $k_T$ of about 30 GeV. 

 In Fig. \ref{fignx-poles} we show the $x$ dependence of the gluon density which includes the  negative $\omega$  contribution (full line) for various $k_T^2$'s. The dashed line shows for comparison the same leading pole contribution as in  Fig. \ref{figx-poles}.  We see clearly from this plot that the negative $\omega$ contribution substantially affects the linearity
in the region of $x \, > \, 10^{-3}$.
Since this non-negligible contribution from  negative $\omega$ occurs at the relatively low  values of $k_T^2$
 considered here whereas data strongly indicates  linear trajectories in $x$ up to $x \sim 10^{-2}$, we take this as a strong
 hint that the infrared phase, $\eta_{NP}$ for $\omega \, < \, 0$ and the precise form of the proton impact factor
 are such that the overlap of the  proton impact factor with the negative $\omega$ part of the Green-function
 is very small. In view of the fact that the proton impact
factor is expected only  to have significant support
near $t_0$ where the infrared phase, $\eta_{NP}$,  is
fixed, such suppression of the overlap would occur if the
infrared phase were very small in the region $\omega \sim 0$.

In this paper we do not attempt to make any detailed comparison with data because we are working here in the LO only and it is well known that the NLO and LO  results 
differ substantially in BFKL. We note, however,  that the difference between the full gluon density and the leading pole contribution seen in Fig. \ref{figx-poles} is due to the subleading poles. A similar change of slope $\lambda$  with $Q^2$~\footnote{ The measured variable $Q^2$ is closely related to the BFKL variable $k_T^2$.} , even more prominent then in our LO calculation, is also seen in the data~\cite{Cald}.  Therefore it is highly possible that the properties of the subleading poles  can be well determined from the
measurements of the slope  $\lambda$ since in our solution there is only one free parameter per pole, $\eta_{NP}$, and the number of contributing poles is small, due to the fast convergence of their sum.  
\section{The Effect of New Physics}
\label{sec9}

\begin{figure}[htbp]
\centerline{\includegraphics[width=12cm]{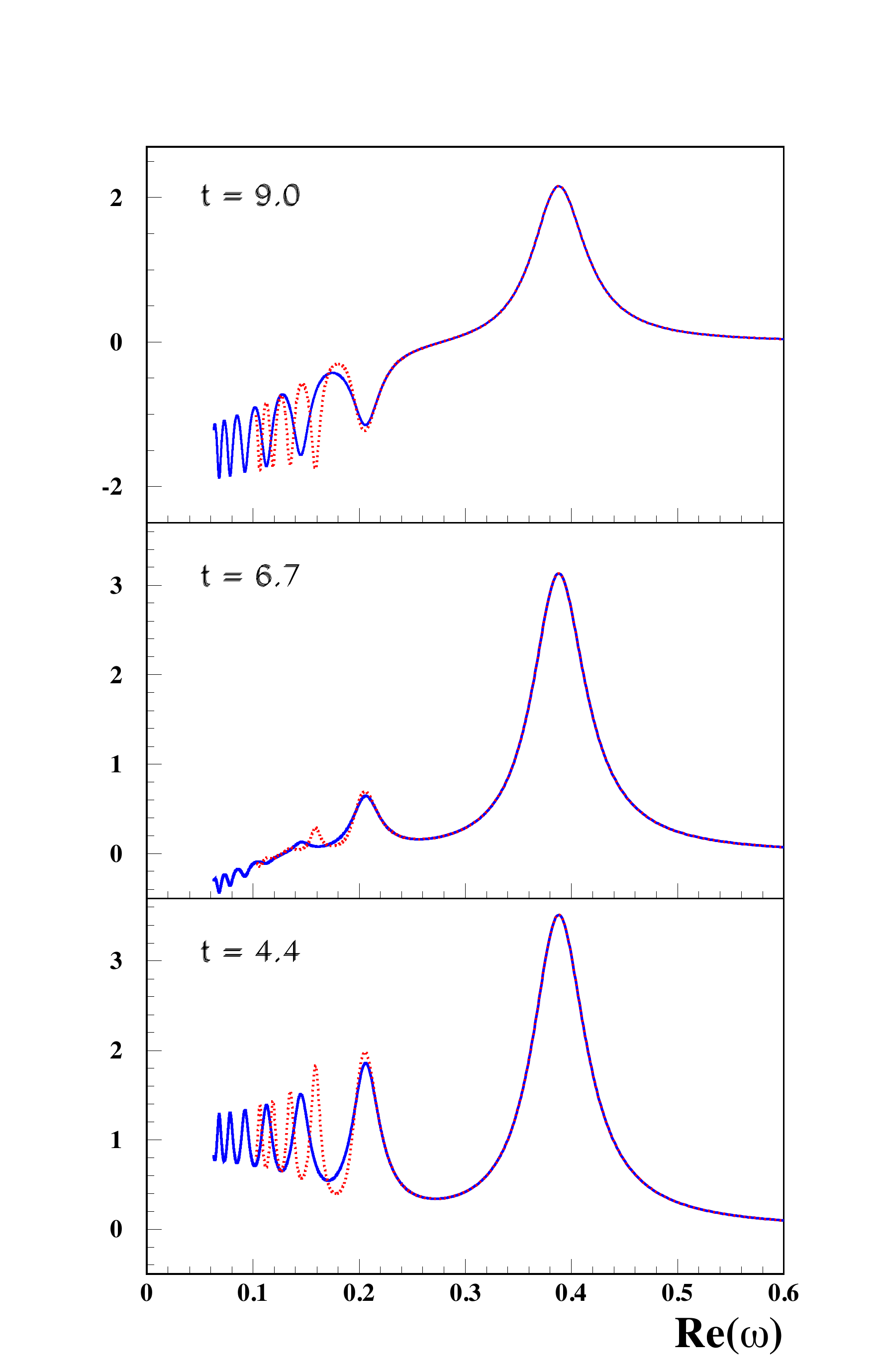} }
\caption{The position of the poles in the Green function for the Standard Model (solid line) and the MSSM with the mass of all super-partners 
taken to be at 3 TeV (dotted line).  We have taken $t=9, \, 6.7, \, 4.4$, corresponding to $k_T$ of 30, 10, 3 GeV,  and $ \ t^\prime=2$.  }
\label{fig11} \end{figure}

In previous publications \cite{KLR,KLRW}, we have pointed out that
 both the positions of the poles in $\omega$ and their residues are sensitive
 to any new physics which affects the running of the coupling and hence 
the value of the critical momentum, $t_c$, provided $t_c$ is above the
 threshold for such new physics. This unique effect is due to the fact that an eigenfunction of the BFKL kernel, (\ref{normeigen}), is a (quasi) bound state of gluons with very different virtualities, ranging from $t_0$ to $t_c$. Even if such a state is probed at low $t$ value, much below the Beyond Standard Model (BSM) threshold, the result is sensitive to the properties of the whole state since the eigenvalue and the residue at the probed $t$ is determined by all the gluons of the state\footnote{Since $t=\ln(k_T^2/\Lambda^2)$, from Table 1 we see that $k_T$  which corresponds to $t_c$ of the first pole is 25 GeV, of the second one is about 4 TeV, of the third one  500 TeV, of the fourth one  54000 TeV, etc. }. 
 
 Formally speaking the sensitivity to BSM thresholds emerge from the fact that the function $\phi(\omega)$,   eq. (\ref{phiom}), which defines the eigenvalues $\omega_n$ by the requirement $\phi(\omega)=n\pi$,  contains an integral over the frequency $\nu_{\omega}(t)$ which ranges from $t_0$ to $t_c$.  
 This frequency, defined by  eq. (\ref{nuom}), is  strongly sensitive to a supersymmetry (SUSY) threshold because the value of $\betabar $ changes substantially in the SUSY region.

 In Fig. \ref{fig11} we show an example of this
 in which we plot the Green function as a function of $\omega$
 on a path close to the real axis, for a typical low $t$ values
 of 9, 6.7 and 4.4 corresponding to $k_T$ of 30, 10 and 3 GeV. We are comparing  the Standard Model  (SM) with the MSSM
 with a supersymmetry (SUSY) threshold of 3 TeV. We see that the position
 of the first  pole is unaffected because the corresponding $t_c$ lies much below
 the SUSY threshold of  3 TeV. However the positions of the sub-leading poles  are shifted to the right because their $t_c$'s are either close to the 3 TeV threshold, 
 as in the case of the second pole,  or much above it for the rest.
  It is interesting to observe that the residues of the non-leading poles oscillates strongly in the displayed $t$ region. This may help to disentangle their contribution since this $t$ region is well accessible to  high precision measurements.

\section{Conclusions and Outlook}
\label{sec10}
We have investigated here the properties of the complete Green function solution to BFKL equation in LO approximation, using a mixed technique of analytic and numerical
 analyses.  We  have shown that this solution fulfills  the completeness requirement and leads to a set of eigenfunctions which are properly normalized and are orthogonal to each other. These  mathematical properties are fulfilled with a high numerical precision, which is not trivial in view of the fact that the eigenfunction states are defined only for $t$ values above some small, but perturbative  $t_0$, and not in the whole region $-\infty < t< \infty$, as would be mathematically required.     
To achieve the completeness it is mandatory to take into account the contribution from the states of the negative $\omega$ continuum. 

The unintegrated gluon density is defined, in this paper, as the inverse Mellin transform for the Green function of the BFKL equation over a suitable path in the complex $\omega$ plane. We show that at low $t$ the integration over an $\omega$ path is exactly equivalent to the sum of poles supplemented by a possible contribution of the negative $\omega$ continuum. 
The sum of poles is dominated by a relatively few terms, which is in contrast to the situation found in the discrete eigenfunction solution of~\cite{KLRW,KLR,lipatov86}. The fast convergence of the complete Green function solution is due to the presence of the $t$ dependent normalization factor which was missing in the pure eigenfunction approach of~\cite{KLRW,KLR,lipatov86}.   			

We have investigated the region of validity of the unintegrated gluon density obtained in this paper and found that it is limited to the region $\Delta t = t - t' < \log(1/x)$. If $\Delta t$ is sizably larger the unintegrated gluon density starts to exhibit an oscillatory behaviour, which eventually leads to an unphysical, i.e. negative gluon density.   Since in DIS $t'$ is limited by the proton factor, which  confines it to very small values, the $t$ values cannot be too large.
  At  $x =10^{-2}$ they  corresponds to $k_T$ values of the order of 10 GeV.   At smaller $x$, like $x =10^{-3}$, the virtuality $k_T$  would be an order or more of magnitude larger, however at  HERA or LHC the region of accessible virtuality decreases with $x$ substantially,  so that the region of applicability of the BFKL equation is limited effectively to $k_T$ of the order of 10 GeV.   

At low $k_T$  the gluon density is dominated by the leading pole, which leads to a linear behaviour in the logarithmic $x$ dependence with a slope which  varies
  with $k_T$. This  variation is due to the contribution of the subleading poles. We show here that the $\omega$ values of the subleading poles are sensitive to the Beyond Standard Model (BSM) effects due to the same mechanism of threshold sensitivity as investigated in  ~\cite{KLRW,KLR}. Therefore, the deviations from the leading pole behaviour are sensitive to the BSM effects and could be measurable due to the high  precision of the
 measured  slopes of the logarithmic $x$ distribution, especially in future projects, see e.g.~\cite{LHeC,Cald-Wing}.   However, in this paper we did not attempt  to perform any data analysis because we are working here in the LO order approximation only and it is well known that the $\omega$ values differ substantially between LO and NLO approximation of BFKL. The LO analysis presented here allows the full understanding of the qualitative properties of the BFKL solution in the low $k_T$ region which we expect to be the same as in the NLO analysis. The quantitative   results in the NLO approximation will be presented in our next paper in which we will also perform a  comparison with the  high precision DIS data.

\section*{Acknowlegments}
We are grateful to Jochen Bartels and Agustin Sabio-Vera for useful
conversations.
One of us (LL) would like to thank  the State University of St. Petersburg for the grant
SPSU 11.38.223.2015  and the grant RFFI 13-02-01246 for support.
One of us (DAR) wishes to thank the Leverhulme Trust for an
Emeritus Fellowship.


\begin{thebibliography}{99}



\bibitem{KLRW}
   H.~Kowalski, L.N.~Lipatov, D.~A.~Ross, and G.~Watt
  Eur.\ Phys.\  J {\bf C70} (2010) 983; \\
 Nucl. Phys {\bf A854} (2011) 45


\bibitem{KLR}  
 H.~Kowalski, L.N.~Lipatov,and  D.~A.~Ross, Phys. \ Part. \ Nucl. 
{\bf 44} (2013) 547 




\bibitem{lipatov86}
  L.~N.~Lipatov,
  Sov.\ Phys.\ JETP {\bf 63} (1986) 904.




\bibitem{LKR1}
 L.~N.~Lipatov, H.~Kowalski, and D.~A.~Ross,
Eur.Phys.J. {\bf C74} (2014) 2919 



\bibitem{ccs} 
M.~Ciafaloni and D.~Colferai, Phys. Lett. {\bf B452} (1999) 372 \\
M.~Ciafaloni, D. ~Colferai and D.P. Salam, Phys. Rev. {\bf D60} (1999) 114036\\
M.~Ciafaloni, D. ~Colferai, D.P. Salam and A. Stasto, Phys. Rev. {\bf D66}
 (2002) 054014


\bibitem{Lipatov14} L.N.~Lipatov, AIP Conf.Proc. {\bf 1654} (2015) 070004 


\bibitem{tw}
R.S. Thorne and C.D. White, Phys. Rev. {\bf D75} (2007) 034005




\bibitem{abf} Altarelli, Ball, Forte
G. Altarelli, R.Ball and S.Forte, Nucl. Phys., {\bf B621} (2002) 359  \\
G. Altarelli, R.Ball and S.Forte, Nucl. Phys., {\bf B674} (2003) 459  \\
G. Altarelli, R.Ball and S.Forte, Nucl. Phys., {\bf B743} (2006) 1  


\bibitem{GP}  Howard Georgi and H. David Politzer,
  Phys. Rev. {\bf D 14}, 1829 (1976)



\bibitem{DGLAP}
V.N. Gribov and L.N. Lipatov,  Sov. Nucl.  Phys.  {\bf 15} (1972) 438\\
G.~Altarelli and G.~Parisi, \ Nucl. Phys. {\bf B126} (1977) 298 \\
Yu. L. ~Dokshitzer,  Sov. Phys. JETP {\bf 46} (1977) 46\\


\bibitem{LLS}
E.~Levin, L.N. Lipatov, and M.~Siddikov,
Phys.Rev. {\bf D89} (2014)  074002 


\bibitem{BFKL}
  I.~I.~Balitsky and L.~N.~Lipatov,
  Sov.\ J.\ Nucl.\ Phys.\  {\bf 28} (1978) 822;
  E.~A.~Kuraev, L.~N.~Lipatov and V.~S.~Fadin,
  Sov.\ Phys.\ JETP {\bf 44} (1976) 443;
 V.~S.~Fadin, E.~A.~Kuraev and L.~N.~Lipatov,
 Phys.\ Lett.\  B {\bf 60} (1975) 50.



\bibitem{salam}  G.~P.~Salam,
  JHEP {\bf 9807} (1998) 019.
 
 

\bibitem{Cald}  A. Caldwell, ``Behavior of $\sigma^{\gamma p}$ at Large Coherence
 Lengths'', axXiv:{\bf 0802.0769} (2008)


\bibitem{LHeC}   A Large Hadron Electron Collider at CERN: Report on the Physics and Design Concepts for Machine and Detector,  arXiv:{\bf 1206.2913} (2012)

\bibitem{Cald-Wing}  A. Caldwell and M. Wing , ``VHEeP: A very high energy electronÐproton collider
based on proton-driven plasma wakefield
acceleration",  axXiv:{\bf 1509.00235} (2015)

\end{thebibliography}
\end{document}